\documentclass[preprint]{aastex631}
\shorttitle{AASTeX v6.31 RS Oph with X-ray gratings}
\shortauthors{Orio et al.}
\graphicspath{{./}{figures/}}

\begin{document}

\title{Shocks in the outflow of the RS Oph 2021 eruption observed with X-ray
 gratings}

\correspondingauthor{Marina Orio}
\email{orio@astro.wisc.edu}

\author[0000-0003-1563-9803]{Marina Orio}
\affiliation{Department of Astronomy, University of Wisconsin 
475 N. Charter Str., Madison, WI, USA}
\affiliation{INAF-Padova, vicolo Osservatorio 5,/122
35122 Padova, Italy.}
\author{E. Behar}
\affiliation{Department of Physics, Technion, 32000 Haifa, Israel}
\author[0000-0002-2647-4373]{G. J. M. Luna}
\affiliation{CONICET-Universidad de Buenos Aires, Instituto de Astronom\'ia y F\'isica del Espacio (IAFE), Av. Inte. G\"uiraldes 2620, C1428ZAA, Buenos Aires, Argentina}
\affiliation{Universidad de Buenos Aires, Facultad de Ciencias Exactas y Naturales, Buenos Aires, Argentina.}
\affiliation{Universidad Nacional de Hurlingham, Av. Gdor. Vergara 2222, Villa Tesei, Buenos Aires, Argentina}
\author{J.J. Drake}
\affiliation{Harvard \& Smithsonian Center for Astrophysics, 60 Garden
  Street, Cambridge, MA 02138, USA }
\author{J. Gallagher}
\affiliation{Department of Astronomy, University of Wisconsin
475 N. Charter Str., Madison, WI, USA}
\author{J. S.\ Nichols}
\affiliation{Harvard \& Smithsonian Center for Astrophysics, 60 Garden
  Street, Cambridge, MA 02138, USA }
\author{J.U. Ness}
\affiliation{European Space Agency (ESA), European Space Astronomy Centre (ESAC), Camino Bajo del Castillo s/n, 28692 Villanueva de la Ca\~nada, Madrid, Spain}
\author{A. Dobrotka}
\affiliation{Advanced Technologies Research Institute, Faculty of Materials Science and Technology in Trnava, Slovak University of Technology in Bratislava, Bottova 25, 917 24 Trnava, Slovakia}
\author{J. Mikolajewska}
\affiliation{Nicolaus Copernicus Astronomical Center, Polish Academy of Sciences, Bartycka 18, 00716 Warsaw, Poland}
\author{M. Della Valle}
\affiliation{Capodimonte Astronomical Observatory, INAF-Napoli, Salita Moiariello 16, I-80131, Napoli, Italy; ICRANet, Piazza della Repubblica 10, I-65122 Pescara, Italy}
\author{R. Ignace}
\affiliation{Physics \& Astronomy, East Tennessee State University, Johnson City, TN 37615, USA}
\author{R. Rahin}
\affiliation{Department of Physics, Technion, 32000 Haifa, Israel}


\begin{abstract}
The 2021 outburst of the symbiotic recurrent nova RS Oph was observed
 with the {\sl Chandra High Energy Transmission
 Gratings} (HETG) on day 18 after optical maximum and with {\sl XMM-Newton}
 and its {\sl Reflection Grating Spectrographs} (RGS) on day 21, before
 the supersoft X-ray source emerged and when the emission was
 due to shocked ejecta. The absorbed flux in the HETG 1.3-31 \AA 
 \ range was 2.6 $\times 10^{-10}$ erg cm$^{-2}$ s$^{-1}$,
three orders of magnitude lower than the $\gamma$-ray flux measured
 on the same date.  The spectra are 
 well fitted with two components of thermal plasma in collisional
 ionization equilibrium, one at a temperature $\simeq$0.75 keV,
 and the other at temperature in the  2.5-3.4 keV range.  
 With the RGS we
 measured an average flux  1.53 $\times 10^{-10}$ erg cm$^{-2}$ s$^{-1}$
 in the 5-35 \AA \ range, but the flux in the continuum and especially in 
 the lines in the 23-35 \AA \ range decreased
 during the 50 ks RGS exposure by almost 10\%, indicating
 short term variability on hours' time scale.
The RGS spectrum  can be fitted with three thermal components,
 respectively at 
 plasma temperature between 70 and 150 eV, 0.64 keV and  2.4 keV.
 The post-maximum epochs of the
 exposures fall between those of two grating
 spectra observed in the 2006 eruption on days 14 and 26:
 they are consistent with a similar spectral evolution, but
 in 2021 cooling seems to have been more rapid. 
  Iron is depleted in the ejecta with respect to solar values, while 
 nitrogen is enhanced. 
\end{abstract}
\keywords{Classical Novae (251);  stars: individual: RS Oph; High energy astrophysics(739); X-rays: stars  }


\section{Introduction \label{sec:intro}}
RS Oph is arguably the best known recurrent symbiotic nova. Classical
 and recurrent novae are binary systems hosting a white dwarf (WD), and their outbursts
 are attributed to a thermonuclear runaway (TNR) on the surface of the
 WD that is accreting material from its binary companion. The TNR is 
 usually followed by a radiation-driven wind, that depletes the
 accreted envelope \citep{Starrfield2012,
 Wolf2013}. The designation ``recurrent'' implies that the
 outburst is observed more than once over human life timescales, 
 although all novae (including the {\it classical}) are thought to be
 recurrent, on secular timescales that can greatly vary, depending
 on the mass accretion rate and the WD mass. The more massive the WD
 is, the smaller its radius, so the accumulated material is more
 degenerate and is ignited with lower accreted mass \citep{Yaron2005,
 Starrfield2012, Wolf2013}. Thus, the frequently
 erupting recurrent novae host rather
 massive WDs. RS~Oph is also a symbiotic system, that is a system
 with a red giant companion, specifically a M0-2 III mass donor
\citep{Dobrzycka1996, Anupama1999} in a
binary with a 453.6 day orbital period \citep{Brandi2009}.
\citet{Brandi2009, Miko2017} presented
 compelling evidence that the WD is very massive, in the 1.2-1.4 M$_\odot$
 range, and that it is a carbon-oxygen WD. The effective temperature
 estimated in the supersoft X-ray phase by \citet{Nelson2008}
 in 2006 was about 800,000 K, which is indicative of a mass
 of at least 1.2 M$_\odot$.
 This implies that it must have grown in mass
 and not have ejected all the accreted material, since 
 calculations indicate that the 
 mass of newly formed WDs is $<$ 1.2 M$_\odot$ , in most models 
 not exceeding 1.1 M$_\odot$ 
\citep[see, among others,][]{Meng2008}.  Such a high CO WD mass
 has spurred much interest in the possibility that RS Oph
 is a type Ia supernova progenitor. 
   RS Oph was observed in outburst in 1898, 1933, 1958, 1967, 1985, and 2006. 
  Two additional outbursts may have been missed in  1907 and 1945 when
  RS Oph was aligned with the Sun \citep{Schaefer2004}.
 
 Novae are luminous at all wavelengths from
 gamma-rays to radio, and X-rays have proven to be a very important 
 window to understanding their physics since the '80ies \citep{Ogelman1984}.
 After the thermonuclear flash, the WD atmosphere contracts and returns
 almost to pre-outburst radius. The peak wavelength of the emission
 moves from the optical range to the UV and extreme UV and finally,
 soft X-rays within a short time, in a phase of constant
 bolometric luminosity, still powered by shell burning. The 
 central source appears as a supersoft X-ray source, with peak temperature
 up to a million K, for a time lasting from days to
 a few years \citep{Orio2012}.
 The first outburst that could be monitored in X-rays was the one 
 of 2006, with {\sl Swift}, RXTE 
\citep{Bode2006, Hachisu2007, Osborne2011,Sokoloski2006},
 and with high spectral resolution with the gratings of
 {\sl Chandra} and {\sl XMM-Newton} \citep{Ness2007,
 Nelson2008, Drake2009, Ness2009}. 
 In outburst, RS Oph is one the most luminous novae in X-rays,
 even if several novae have been observed
 by now much closer to us. The distance inferred from the expansion
 velocity and resolved radio imaging of
 2006 is 2.45$\pm$0.37 kpc \citep{Rupen2008},
 Assuming that the giant fills its Roche Lobe, and using the
orbital parameters \citet{Brandi2009}, the resulting distance is
3.1+/-0.5 kpc \citep{Barry2008}.  Finally,the GAIA DR-3 distance
 is 2.44$^{+0.08}_{-0.16}$ kpc (geometric) and 2.44$^{+0.21}_{-0.22}$
 kpc (photogeometric) \citep[see][]{Bailer-Jones2021}.
 The uncertainty on the GAIA parallax may be larger than estimated,
 because of difficulties arising from the surrounding nebula  and the
 wobble of the long binary period, however the derived
 distance range is in good agreement
 with the most recent and updated measurements.  
Some of the older estimates of the distance to RS Oph 
 are no longer relevant now, since the binary parameters are better known.
 We note that many papers in the past adopted the 1987 estimate of 1.6 kpc based
 on the intervening neutral hydrogen column density to the source
\citep{Bode1987, Hjellming1986}. 

 RS Oph's latest outburst was observed on August 9 2021 22:20 UT 
 (as announced in \url{http://ooruri.kusastro.kyoto-u.ac.jp/mailarchive/vsnet-alert/26131}) and \url{http://www.cbat.eps.harvard.edu/iau/cbet/005000/CBET005013.txt} at visual magnitude 4.8. 
 Immediately afterwards, the nova was also detected at gamma-ray energy with 
 the Fermi-LAT \citep{Cheung2022}, H.E.S.S. \citep{Wagner2021,
 Wagner2021b, HESS2022} and MAGIC \citep{Acciari2022} , and in
 hard X-rays with MAXI \citep{Shidatsu2021}, and INTEGRAL \citep{Ferrigno2021}. 
 The highest energy range was that of H.E.S.S. and
 MAGIC, from 10 GeV to tens of TeV,
 and the flux peaked in its ``softest'' range, with a power law index $>$3,
 significantly higher than the 1.9 power law index
 in the spectrum measured with the Fermi-LAT in the 100 MeV-13 GeV range.

 The early optical spectrum was described by \citet{Munari2021}
 as of ``He/N'' type, with strong Balmer, He I and N II lines. The
 emission lines had full width at half maximum 2900 km s$^{-1}$ and blue 
 shifted P-Cyg components, that disappeared within a few days 
 \citep{Miko2021, Munari2021b}.
 Acceleration to up to $\simeq$4700 km s$^{-1}$ was observed after
 the first two days, and P-Cyg profiles appeared also in lines of Fe II, O I, and Mg II \citep{Miko2021, Pandey2022}.
 A narrow emission component disappeared within the first few days,
 while
 a narrow absorption component persisted for longer \citep{Luna2021, Shore2021}, and altogether
 the velocity of the lines indicated deceleration \citep{Munari2021b}
 a few days after the initial acceleration.
 Intrinsic linear optical polarization was observed $\sim$1.9 days after outburst \citep{Nikolov2021} while satellite components appeared in the optical spectra after two weeks
 in H $\alpha$ and H $\beta$, suggesting a bipolar outflow as observed
 in the radio in 2006 \citep{Rupen2008}. High ionization lines
 appeared around day 18 of the outburst \citep{Shore2021}. A summary
 and visual illustration of  the optical spectral changes in the  
 first 3 weeks after maximum can be found in \citet{Munari2021c, Pandey2022}.

 The AAVSO
 optical light curve of RS~Oph in different bands, from B to I, in 2021
appeared
 extremely similar to the AAVSO 2006 light curve. 
 The maximum magnitude was V=4.8, the time for a decay by 2 magnitudes t$_2$
 was 7 days and the time for a decay by 3 magnitudes t$_3$ was 14 days. 
 All the subsequent evolution was smooth, and in the last optical observations
 on November 14 2021 the nova was at V$\simeq$11.2, like in 2006 at the
 same post-outburst epoch. 
 However, the X-ray light curve was quite different from the 2006 one,
 with a shorter and less luminous supersoft X-ray phase
 \citep{Page2022}, and Orio et al. (2022 article in preparation on
 {\sl NICER} monitoring).

 Here we 
 analyze and discuss high-resolution X-ray spectroscopy obtained
 with the {\sl Chandra} HETG) on 2021-8-27 and
 with the {\sl XMM-Newton} RGS on 2021-8-30. 
 Section 2 describes the Chandra spectrum, model fitting and lines
 profiles and fluxes.
 Section 3 describes the XMM-Newton spectra and light curves, again
 with spectral fits and lines' fluxes and profiles. Section 4
 describes a comparison with observations done during the 2006 outburst,
 Section 5 is dedicated to a discussion of our findings and Conclusions
 are in Section 6.
\section{The Chandra observation}
 The HEG (high energy grating) and MEG (medium energy grating)
 were coupled with the ACIS-S detector
 and spectra were obtained on 2021-8-27 in a 20~ks long exposure 
starting at 00:56:43.
We extracted these spectra with the
CIAO software \citep{Fruscione2006} version 4.14.0 and the CALDB
calibration package version 4.9.7.
Both the HEG and MEG were used, 
with a respective absolute wavelength accuracy of 0.0006 \AA \
and 0.011 \AA. We measured a count rate 0.444$\pm$0.004 cts s$^{-1}$
 in the zeroth order ACIS-S camera 
 (with an estimated pile-up fraction of about 10\%).
We extracted the ACIS-S lightcurve and found no significant variability
 during the exposure.

 Half of the incident radiation is 
dispersed to the gratings in this observation
mode, and the count rates were 1.622$\pm$0.010 cts s$^{-1}$ in the HEG +1/-1
 orders  (energy range 0.8-10 keV, wavelength range 1.2-15 \AA),
and  1.271$\pm$0.013 s$^{-1}$ in the MEG summed +1/-1 orders
 (energy range 0.4-5 keV, wavelength range 2.5-31 \AA). The
 integrated fluxes in the two instruments were measured as 
 2.367 $\times 10^{-10}$ erg cm$^{-2}$ s$^{-1}$ in the HEG
 and 2.605 $\times 10^{-10}$ erg cm$^{-2}$ s$^{-1}$ in the MEG.
 The HEG and MEG spectra, shown in Figure~1, are rich in prominent emission lines due
 to H-like and He-like transitions, from calcium to neon.

\subsection{Model fitting to the HETG spectra}
Spectral fitting was done with the HEASOFT XSPEC tool, version 6.30.1,
 after the data were binned by
signal-to-noise with the GRPPHA tool 
\citep[see][and references therein]{Dorman2001}.

 Figure~1 shows in red the best fit
 with a 2-component BVAPEC model in XSPEC, of thermal plasma 
in collisional
 ionization equilibrium with line broadening and variable abundances,
 and a TBABS model of absorbing column density.
 With $\chi^2$ statistics, which we found to be appropriate in this
 case, we obtained a value of $\chi^2$ per degrees of freedom 
 ( $\chi^2$/d.o.f.) of about 1. The parameters of the fits are given in Table 1.
 In addition to the line broadening, the global fit is improved by allowing
 a blueshift as free parameter. This blueshift velocity necessary
 in the fit is modest with respect
 to typical nova wind velocity, only 264 km s$^{-1}$, and 
 in previous X-ray observations of novae it has been explained as 
 an apparent effect of differential absorption, eroding the red wing of each
 line more than the blue wing because material moving towards us is
 absorbed by fewer layers of intrinsic absorbing material than the
 receding outflow \citep[see]{2001ApJ...549L.119I, Orlando2009, Drake2009, Drake2016, Orlando2017, Orio2020}. 
 We explore this effect more in the next section.
 
 We allowed the abundances of elements from neon to iron
 that have many transitions in the observed energy range, 
 (specifically Ne, Mg, Al, Si, S, Ar, Ca and Fe)  to be free parameters,
 but, in order to reduce the number of free parameters,
  we assumed these abundances to
   be the same in the two components, although this may not be so 
 \citep[see models by][]{Orlando2009}.   

Figure 2 shows two of the He-like triplets observed
 with the HEG, namely the prominent 
 \ion{Si}{13} and \ion{S}{15} lines. With a Gaussian fit to the lines, 
 for \ion{Si}{13} we found that the G ratio of the sum of the flux in the intercombination and forbidden lines to the
 flux in the resonance line, indicated as $G=(i+f)/r$, is 1.26$\pm$0.12 for
 S XV, 0.82$\pm$0.10 for Si XIII, and 0.77$\pm$0.18 for Mg XI (see Table 2).

 As long as the forbidden line is formed in a certain electron
 density range, which
 depends on the specific element, a value G$<$4 indicates
 that the contribution of photoionization is not important
\citep{Porquet2010}. We discuss 
 line profiles and fluxes more in detail in the next Section, in which we
 also analyse diagnostics of electron density. 
 Figure~3 (where only the model is plotted) shows the respective 
contributions of the two thermal components in
 the model to the continuum, H-like
 and He-like lines. While both components contribute in the same way to
 the continuum and quite similarly to the H-like lines at shorter
 wavelengths, the hot component contributes increasingly less to H-like lines
 formed at lower
 energy. Moreover, it turns out that the gas of the hotter component is
 close to being fully ionized, so its contribution to the He-like lines is
 negligible. 

 In Table 1 we give the 90\% confidence level error for each free parameter
 obtained by assuming the other parameters are fixed.
 We also analyzed the
 confidence contours obtained by varying two parameters simultaneously. 
 We found that the temperature of the softer components varies
 in a narrow range (0.72-0.78 keV within a 2 $\sigma$ uncertainty) if
 we let either N(H) or the temperature of the other component vary. 
 The column density N(H) and the temperature of the hotter
 component are less well constrained. T$_2$ is in the 2.45-3.45 keV
 range within 2 $\sigma$. 
\begin{figure}
\begin{center}
\includegraphics[width=87mm]{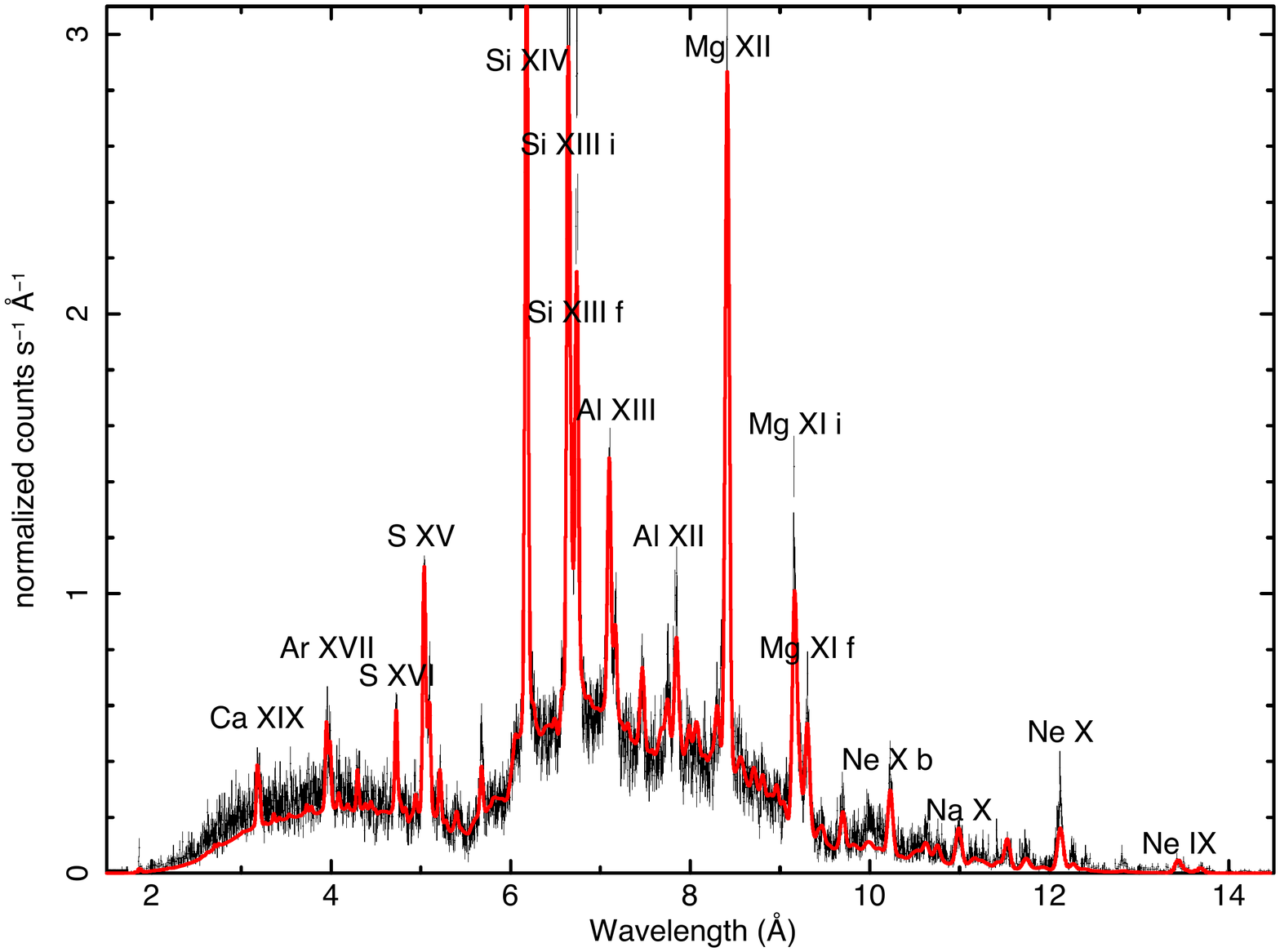}
\includegraphics[width=87mm]{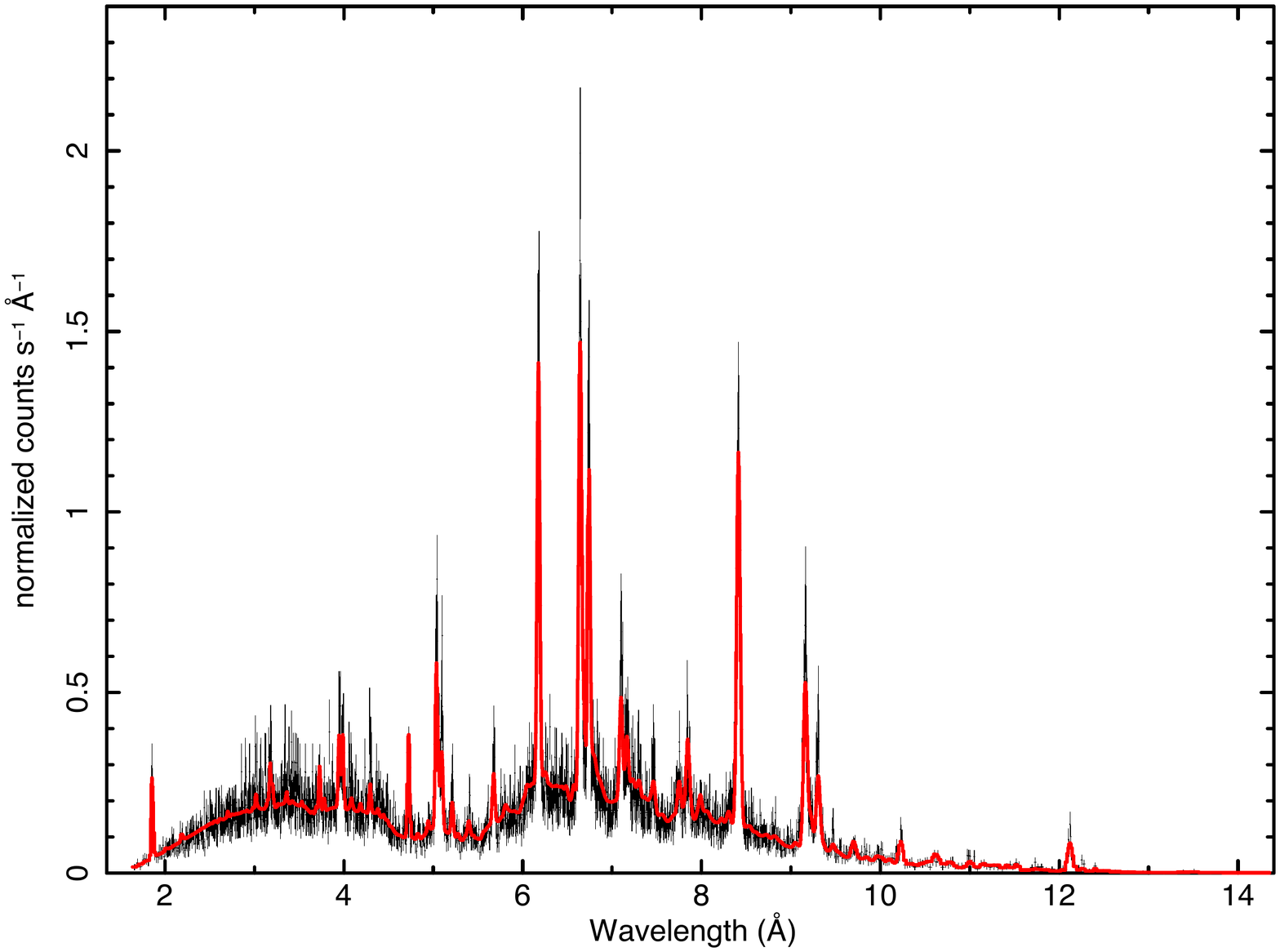}
\end{center}
\caption{The spectrum observed with the  Chandra MEG grating (left) and
 with the Chandra HEG grating (right) on 8-27-2021, day
 18 after optical maximum, and a fit with the 2
 APEC components described in the text, whose parameters are given on Table 1,
 second column.}
\end{figure}
\begin{figure}
\begin{center}
\includegraphics[width=87mm]{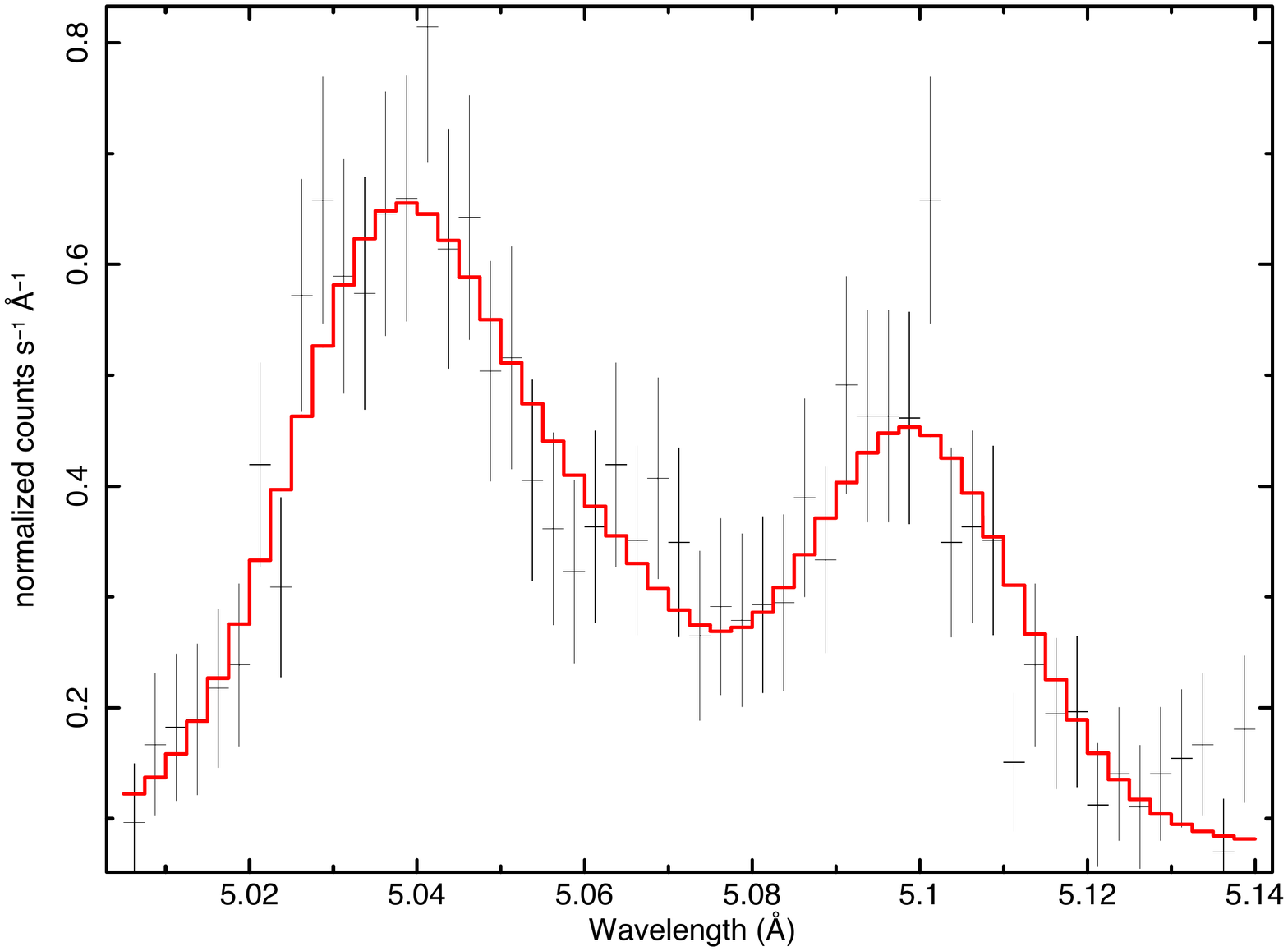}
\includegraphics[width=87mm]{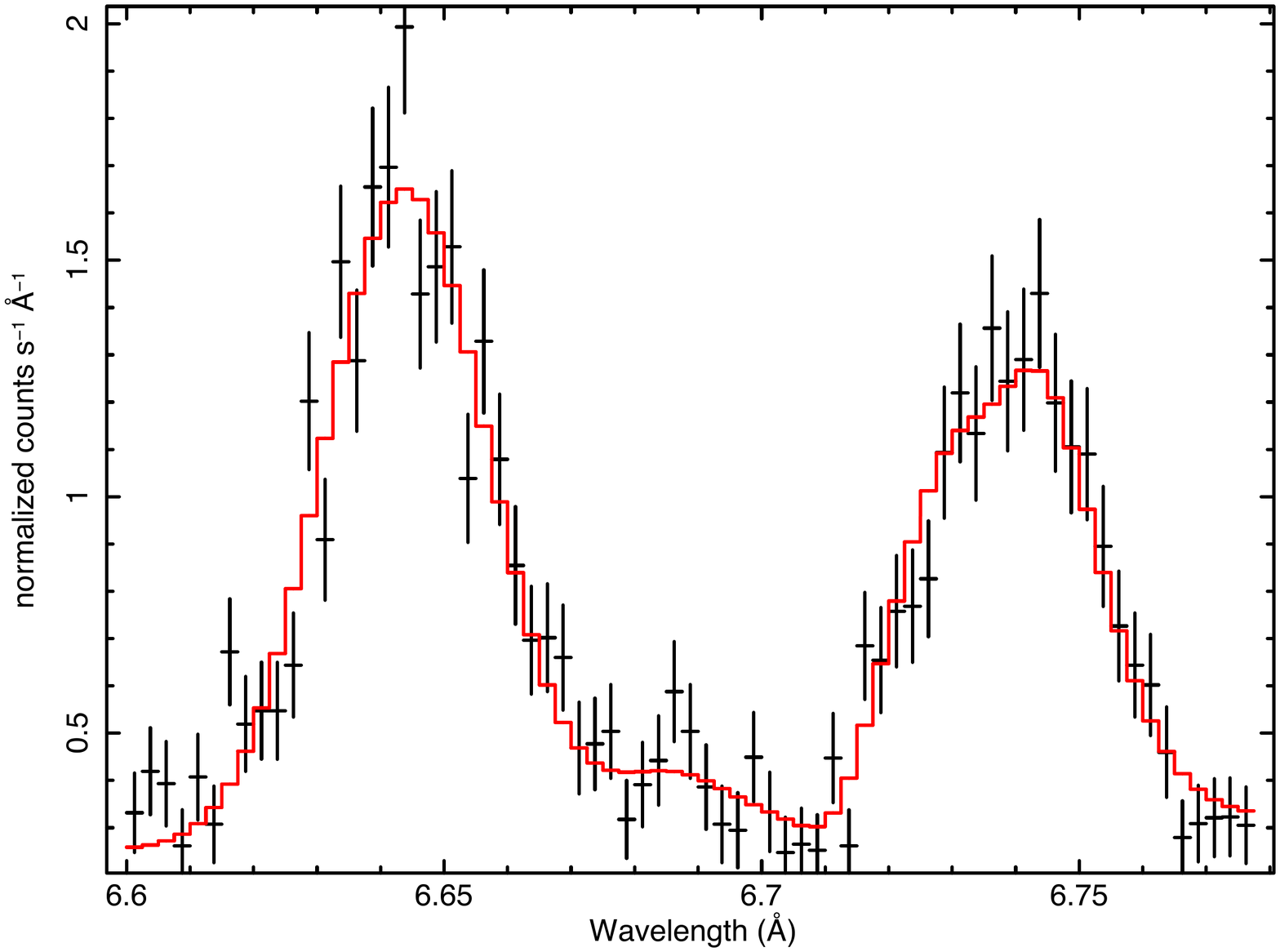}
\end{center}
\caption{The observed profile, in the count rate spectrum, of the Si XIII 
(right)
 S XV (left) He-like triplets observed
 with the HEG. The clear line on the left
 with the highest count rate is the resonance (r), the
 clear line of the right is the forbidden (f), while in the middle an
 intercombination (i) line is barely resolved, 
 The i line is composed of two 
 very close lines in wavelength, hardly resolvable. 
The red lines show the fits with 3 Gaussians with the same width, and 
a locally evaluated underlying continuum, convolved with the instrumental
 response.}
\end{figure}
\begin{figure}
\begin{center}
\includegraphics[width=87mm]{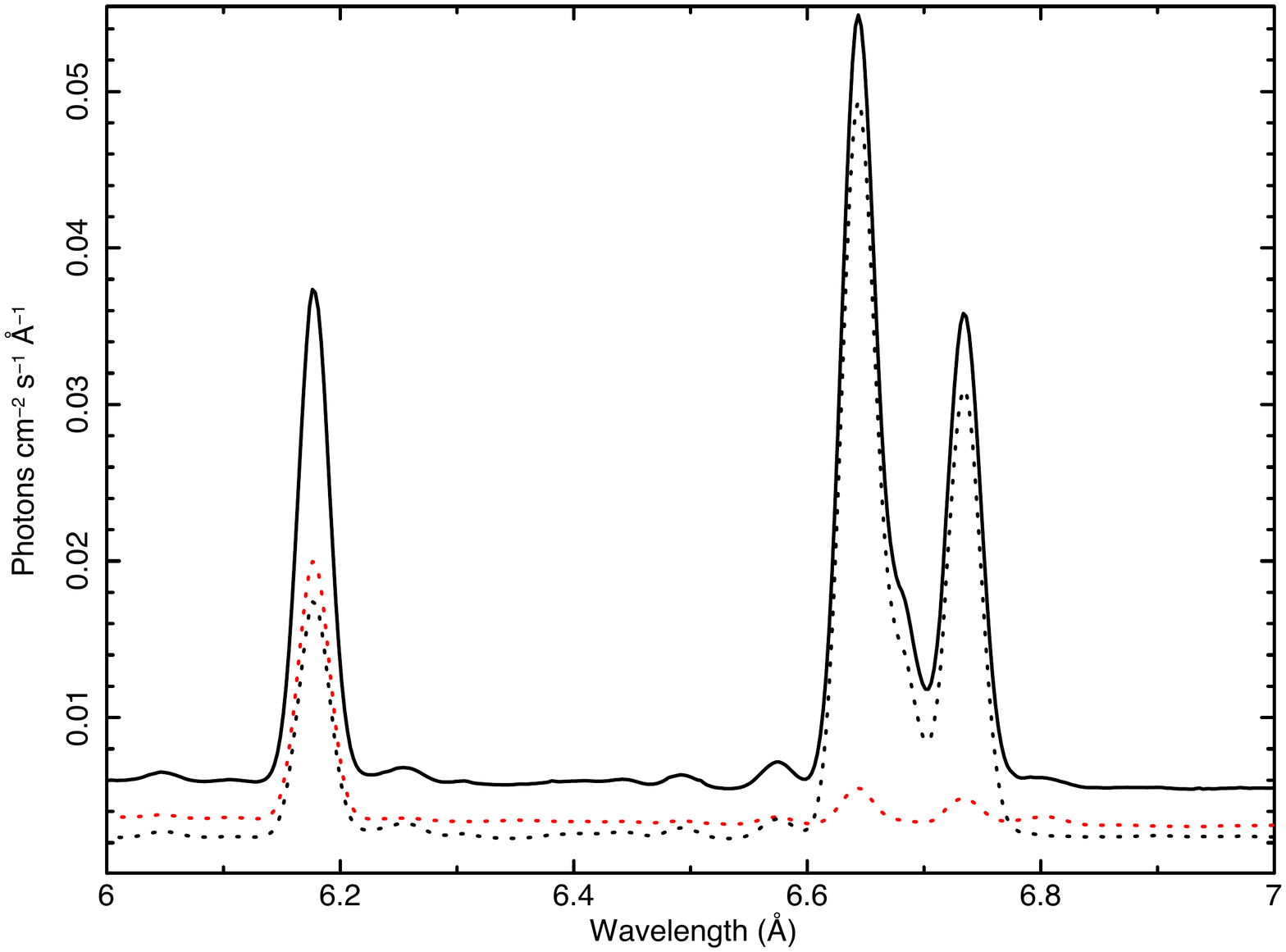}
\includegraphics[width=87mm]{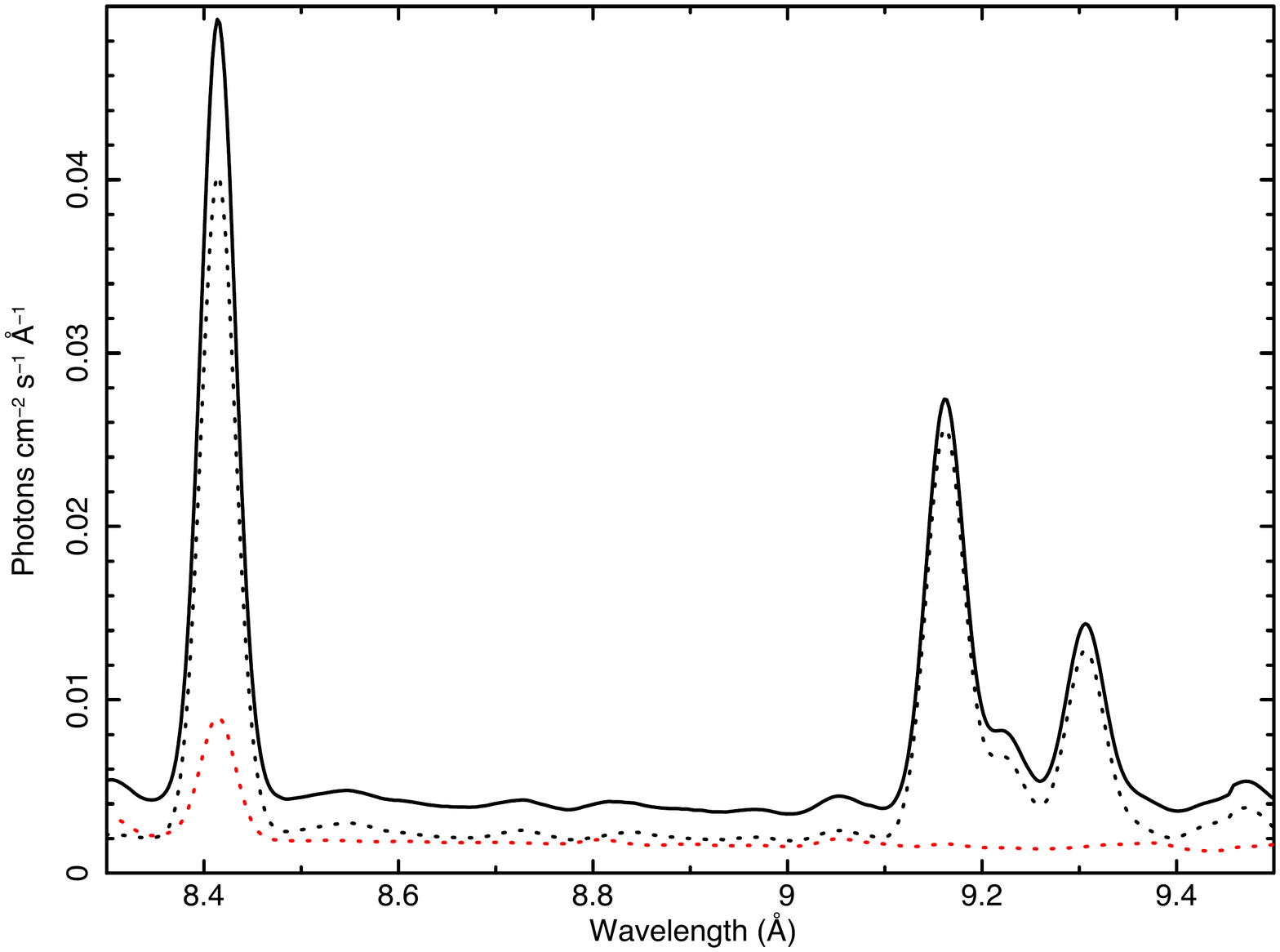}
\end{center}
\caption{A close-up showing the model 
 with two thermal components in Table 1 (left column)
 and in Fig. 1, resulting
 from  fitting the binned spectrum
 with S/N=10 and convolving it with
 the instrument response for the Si XIV H-like, S XIII He-like,
 Mg XII H-like and Mg XI He-like lines. 
 The hotter component is shown in red, the less hot with the black dotted line,
 and the sum of the two with the black solid line.
 As shown also by Fig. 2,
 the intercombination lines are predicted to be non-resolvable, so we see only two clear lines (r and f) in the model.
 }
\end{figure}
%
%
\begin{figure}
\begin{center}
\includegraphics[width=130mm]{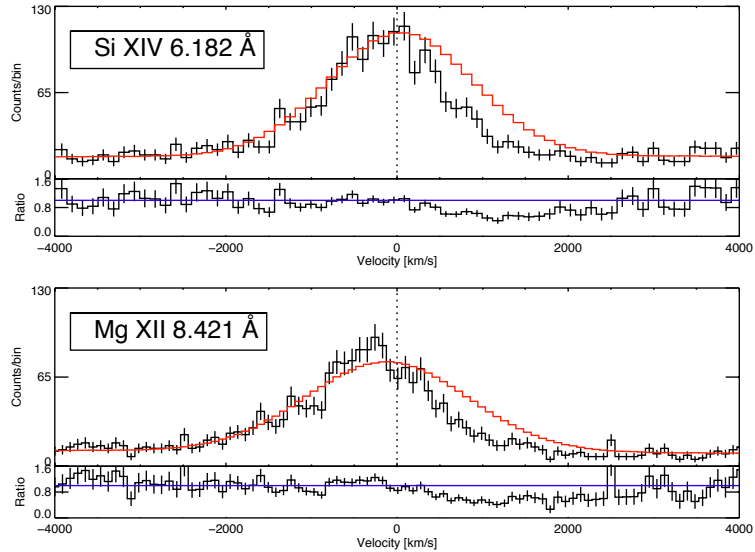}
\end{center}
\caption{HEG line profiles of the 
 the H-like lines of Si XIV  and Mg XII  and fit with
 a Gaussian, in velocity space. The profile is clearly
 asymmetric; we interpret this as the effect of differential
 photoelectric absorption by the outflowing material of the nova, that
 erodes the red wing much more than the  blue one.}
\end{figure}
\begin{table}
\caption{Parameters for the two HETG models described in the text, both with
 two BVAPEC components. The emission measure is the integral
 over the emitting volume of the electron density multiplied by the ion
 density. Assuming electron and ion densities to be 
 equal, so that, in first approximation,
 EM= V$ \times n_{\rm e}^2$. The values are normalized assuming a distance of  2.4 kpc.
Abundances of elements with relevant emission lines are also given, 
 and so are the blue shift and broadening velocity. The statistical uncertainty
 listed  is the largest of the positive and negative
 90\% confidence level.}
\begin{center}
\begin{tabular}{ccc}
\hline
 & N(H) free &  N(H) fixed \\
  & & \\
\hline
 $\chi^2$/d.o.f. & 1.0   & 1.1 \\
 N(H) $\times 10^{21}$ cm$^{-2}$ & 10.3$\pm$0.4 & 5.2  \\ 
 kT$_{1}$ (keV) & 0.76$\pm$0.01 & 0.77$\pm$0.12  \\
 kT$_{2}$ (keV) & 2.83$\pm$0.25 & 3.58 $\pm$0.11  \\  
 EM$_{1} \times (d(Kpc)/2.4)^2$  & (2.835$\pm0.105) \times 10^{58}$ & 
 (1.366$\pm0.038) \times 10^{58}$  \\ 
 EM$_{2} \times (d(Kpc)/2.4)^2$  & (1.574$\pm0.091) \times 10^{58}$  & 
  (1.428$\pm0.033) \times 10^{58}$  \\
 F$_{a,tot}$ $\times 10^{-10}$ erg cm$^{-2}$ s$^{-1}$ & 2.63$\pm$0.15 & 2.78$\pm$0.07 \\
 F$_{a,1}$ $\times 10^{-10}$ erg cm$^{-2}$ s$^{-1}$ & 1.00$\pm$0.04  & 0.84$\pm$0.02 \\
F$_{a,2}$ $\times 10^{-10}$ erg cm$^{-2}$ s$^{-1}$ & 1.63$\pm$0.09 & 1.94$\pm$0.05 \\
F$_{un}$  $\times 10^{-9}$ erg cm$^{-2}$ s$^{-1}$ & 1.05$\pm$0.06 & 0.58$\pm$0.03\\
 Ne/Ne$_\odot$ & 1.15$\pm$0.11 & 0.77$\pm$0.07   \\
 Mg/Mg$_\odot$ & 1.02$\pm$0.04  &  1.21$\pm$0.05  \\
 Al/Al$_\odot$ & 0.92$\pm$0.11  &  1.39$\pm$0.15  \\
 Si/Si$_\odot$ & 0.75$\pm$0.03  &  1.16$\pm$0.05  \\
 S/S$_\odot$  & 0.98$\pm$0.06 &  1.58 $\pm$0.08 \\
 Ar/Ar$_\odot$ & 0.93$\pm$0.12  & 1.23$\pm$0.18   \\
 Ca/Ca$_\odot$ & 0.62$\pm$0.15  & 0.61$\pm$0.20   \\
 Fe/Fe$_\odot$ & 0.28$\pm$0.03  & 0.07$\pm$0.02   \\
 v$_{bs}$ km s$^{-1}$ & 265$\pm$13 & 264$\pm$13  \\
 v$_{br}$ km s$^{-1}$ & 677$\pm$14 & 680$\pm$14 \\ 
\hline
\end{tabular}
\end{center}
\end{table}
 In Table 1 we give both the parameters of the fit that yielded a reduced  
 $\chi^2$/d.o.f.$\simeq 1$ 
 that gave the best fit to the HETG alone, and the fit
 with N(H) constrained by fitting also the spectrum
 obtained from a public {\sl NICER} observation
 done on 2021-08-26 starting at 07:41:00, in an uninterrupted 2170 s
 exposure.  With a joint fit, we obtained a best-fit
 value N(H)=5.2 $\times 10^{21}$ cm$^{-2}$. This
 is still consistent with fitting the HETG spectrum alone,
 because the fits have
 different minima of the  $\chi^2$ parameter
 when we explore  a large parameter space,
 even if the other parameters are constrained within
 quite narrow ranges. The 
 {\sl NICER} monitoring of RS Oph and its results are described in
 an accompanying paper (Orio et al. 2022, in preparation). 

In Table 1, second column, we also list the parameters of  
 the fit to the HETG spectra  fixing
 the value of N(H) previously obtained including {\sl NICER} in the fit. Adding the {\sl
 NICER}  
 broad band spectrum constrained N(H), but introduced larger
 errors in the other parameters, so we show only the values obtained for the HETG,
 fixing the N(H) parameter.
 The new value of N(H), 5.2 $\times 10^{21}$ cm$^{-2}$,
 differs by slightly more than 2 $\sigma$ 
 from the value obtained with the HETG alone, 
 but {\sl NICER} is sensitive to energy as low as 
 0.2 keV, and we inferred that
 the {\sl NICER} fit gives a more precise estimate of N(H). Fixing this 
 N(H) value, we could still 
 fit the HETG spectra with $\chi$/d/o.f. $\simeq 1.1$. 
 Table 1 shows that with this N(H) value, the HETG spectra are fitted with 
almost unvaried temperature of
 the cooler component, but with a higher temperature of the hotter one
 than in the ``free N(H)'' fit. We reasoned that,    
 while the absorption is better
 determined in the lower energy range of {\sl NICER},
 the temperatures of both components are much more reliably estimated
 with
 the HETG because they depend on line intensities and line ratios.
 Both model fits in Table 1 predict sub-solar abundances of calcium and iron,
 moderately enhanced neon and magnesium, while the abundances
 of other elements are not well constrained: they are slightly  
 enhanced in the fit with higher N(H) value, and slightly depleted in the
 fit with N(H) value that also  fits the {\sl NICER} spectrum.

 We tried a third fit, not shown in the Table,
  without ``tying'' the abundances of
 the two components, namely assuming that the abundances are {\it not} the same
 in the two plasma components. With the fixed value of N(H) adopted in
 the second column of Table 1  and also with a free N(H) value, we 
 obtained a value of $\chi^2$/d.o.f.=1 with depleted iron,
 but almost solar abundances of elements from neon to calcium in the 
 hotter component, and largely enhanced ($>$30 times solar) 
 abundances in the less hot one. With the fixed value of N(H), in this model the best fit
 temperatures are slightly different, 2.84 keV and 0.59 keV. The emission lines appear to be 
 somewhat better fitted, but statistically the result is not preferred and
 it has too many free parameters to constrain the abundances
 very significantly.
\subsection {Line profiles and line fluxes}

Table 2 shows the measurements of all the lines that had a significant
 detection above the continuum, fitted with Gaussian profiles
 above a continuum that is estimated locally, in a narrow
 range around each line, with a power law. For the He-like triplets,
 we assumed that the $\sigma$ of the Gaussian (which we attribute
 to broadening velocity and as such is indicated in the Table), is
 the same for the resonance, intercombination and forbidden line. 
 Fig. 2 shows examples of this fit for Si XIII and S XIV.
 We also examined the line variability during the exposure by dividing
 the spectrum into 6 sub-intervals of equal duration,
 and did not measure significant variability in any line fluxes.

The fit with the two thermal components in Table 1 includes a blue-shift
 velocity, corresponding to $\simeq$263$\pm$13 km s$^{-1}$,
 and  line broadening velocity, 667$\pm$14 km s$^{-1}$. In order to assess the
 significance of the line broadening, we compared the line widths 
 with a reference source, TW Hya. We know from
 previous work that TW Hya, compared to
 the ``Rosetta-stone'' Capella,  does not show line broadening 
 exceeding the instrumental width of the line \citep{Brickhouse2010}.
 We thus found that the width of the lines in both HEG and MEG
 is larger than instrumental.

 The blue shift, however, does not necessarily imply a velocity in
 a direction receding from us. As in the previous outburst of RS Oph 
 \citep{Nelson2008, Drake2009}, and in two other symbiotic recurrent novae
 \citep{Drake2016, Orio2020}, the line profiles are non symmetric and skewed towards
 the blue. This is likely to result in an apparent blue-shift
 velocity when fitting the lines with Gaussians or other symmetric
 profiles, and it may be 
 due only to intrinsic differential absorption in the nova ejecta,
 which is greater in the material receding from
 us than in     the material moving towards
 us. Figure 4 illustrates the profiles of the H-like lines of Si XIV
and Mg XII.
 In general, lines at higher wavelength, originating mostly in plasma at lower 
 temperature because they are formed with a lower ionization potential,  
 appear to be more skewed. We show in the next Section
  that 3 days later this effect was quite evident, having observed
 and measured well lines in a ``softer'' range.

\begin{deluxetable}{lrRRRR}
 	 \tablecaption{Line measurements for the HEG and MEG spectrum - the MEG 
 was used for the S XVI, Ne XI and Mg XI lines, the HEG for the rest. The
 whole  Fe XXVI triplet was treated as one Gaussian to estimate
 the flux, because of resolution difficulties. ``fixed'' means that the 
 Gaussian center was evaluated bye eye instead of being a free
 parameter. The Gaussian width of the triplets was assumed to be
 the same for the r, i and f lines. The uncertainty reported
 in the Table is the largest of the 90\% statistical error in the
 + and - directions.}
 	 \tablehead{
 	 \colhead{Ion}& Grating &
 	 \colhead{$\lambda_0$ (\AA)} &
 	 \colhead{$\lambda$ (\AA)} &
 	 	 \colhead{$F_x$ (10$^{-12}$ erg cm$^{-2}$ s$^{-1}$)} &
  \colhead{$\sigma$ (km s$^{-1}$)}\\
 	 }
 	 \startdata
 Fe XXVI & HEG & 1.850-1.868 & 1.859 ({\rm (fixed)}) & 3.961\pm0.780 &  \\
 S XVI & MEG & 4.7273 &   4.725\pm0.003 & 1.333_{-0.255}^{+1.340} & 1016\pm400 \\
 S XV - r & HEG & 5.0387 & 5.038\pm0.003 &  3.006$\pm$0.451  & 289$\pm$30 \\
 S XV - i &     & 5.0631,5.0616 & 5.068\pm0.002 & 1.326$\pm$0.351 &  \\
 S XV - f &     & 5.102 &  5.099\pm0.0045 & 2.462$\pm$0.391 &  \\
 Si XIV & HEG & 6.180-6.186   & 6.179\pm0.003 & 3.530\pm0.247 & 146\pm15 \\
        & RGS &               & 6.177\pm0.004 & 3.217\pm1.220 & 194\pm20 \\
Si XIII - r & HEG & 6.648 &  6.644\pm0.002 & 4.661\pm0.261   & 267\pm15 \\
            & RGS &       &  6.640 & 4.562  & 1174\pm633 \\
 Si XIII - i &     & 6.685, 6.688 &    6.684\pm0.006 & 0.716\pm0.035   &   \\
             & RGS &              &   6.633  ({\rm (fixed)}) & 0.938\pm0.094  & \\
 Si XIII - f &     & 6.7403 &  6.735\pm0.001 & 3.095\pm0.217 &  \\
             & RGS &        &  6.723 {\rm (fixed)} & 3.769\pm0.754  & \\ 
 Mg XII & HEG & 8.419 &  8.411\pm0.003 & 4.172\pm0.350 & 463\pm0.48  \\
        & RGS &       &  8.405\pm0.004 & 5.153\pm0.234 & 999\pm131   \\
Mg XI - r & MEG & 9.169 & 9.157\pm0.028 & 2.764\pm0.195  & 688\pm48 \\
          & RGS &       & 9.156         & 3.387\pm0.256 & 2588\pm439 \\
Mg XI - i & MEG & 9.228 9,231 & 9.219\pm0.004 & 0.798\pm0.123 &  \\
          & RGS &             & 9.205 ({\rm (fixed)}) & 0.964\pm0.193 & \\  
Mg XI - f & MEG    & 9.314 & 9.306\pm0.004 & 1.337\pm0.078 &    \\
          & RGS    &       & 9.294   & 1.975\pm0.197 & \\ 
Ne X &  MEG & 12.132 (12.138) & 12.120 ({\rm (fixed)})
 & 2.068\pm0.168 & 445\pm67 \\ 
     &  RGS & 12.132 (12.138) & 12.112\pm4  & 3.999\pm0.299 & 1804\pm170 \\
Fe XVII & RGS & 15.031 & 15.001 {\rm (fixed)} & 1.669\pm0.073  & 480\pm53 \\
Fe XVII & RGS & 15.261 & 15.232\pm0.024 & 4.874\pm0.561  &  \\
Fe XVIII? & RGS & 15.232 & 15.157 ({\rm (fixed)}) & 0.226\pm0.043  &  \\
N VII & RGS & 24.779, 24.785 & 24.716\pm0.005 & 1.269\pm0.040 & 956\pm37 \\
NVI - r  & RGS & 28.780 & 28.760\pm0.009 & 4.077\pm0.068  & 1313 \\
NVI - i  & RGS & 28.0819, 29.0843 & 29.084\pm0.499 & 0.298\pm0.021 &  \\
NVI - f  & RGS & 29.5346  & 29.449\pm0.016 & 1.416\pm0.014 &   \\
\enddata
\end{deluxetable}

From the fit in Table 1 we have values of the emission measure (assuming
 a distance of 2.4 kpc).  We know that the outflow is mainly on
 the equatorial plane and from the poles, and that the filling factor is low, but
 in a simplistic assumption that all the volume was filled in 18 days by
 material outflowing at 3500 km s$^{-1}$ and adding
 the emission measures of the two components, we obtain an
average value of the electron 
 density of 6.96 x 10$^6$ cm$^{-3}$ for the second model, and a
8.67 x 10$^6$ cm$^{-3}$  for the first. 
 However, the electron density in novae at this stage is
 likely to be higher, around  10$^{9}$ cm$^{-3}$ \citep[e.g.][]{Neff1978}.
 The average value obtained from the emission measure is often much lower
 because of clumping and low filling factor 
 \citep[see discussion by][]{Orio2020}. Low filling factors,
 in the 10$^{-5}$-10$^{-5}$ range, are often found 
 also analysing optical spectra \citep[e.g.][]{Snijders1984, Andrea1991}.

The R=f/i ratio
 obtained from the HETG is 1.86$\pm$0.31 for S XV, 4.32$\pm$0.09 for Si XIII,
 and 1.68$\pm$0.12 for Mg XI.
 These ratios can be compared with Fig. 6 in \citet{Orio2020}, showing the f
 ratio as a function of density in absence of strong photo-excitation in 
 three models in the literature. From
 this comparison, we find that the R value estimated
  for S XV and Si XIII  do not constrain the electron density,
 but the value obtained for
 S XV gives an upper limit of about 10$^{14}$ cm$^{-3}$. 
 The value for Mg XI indicates about 3.16 $\times 10^{12}$ cm$^{-3}$
 and a {\it lower} limit of 2 $\times  10^{13}$ cm$^{-3}$,
 which does indicate very significant clumping when compared with
 the average density obtained from the emission measure. 
\section{The XMM-Newton observation}
 A second X-ray grating spectrum was obtained on August 30 2021 in 
 a 54 ks continuous {\sl XMM-Newton exposure} starting at UT 15:45:41. 
 The {\sl XMM-Newton} observatory consists of five 
instruments behind three X-ray mirrors, plus an optical
monitor (OM); all instruments observe simultaneously. For this paper, we
 used the spectra of the Reflection Grating Spectrometers 
(RGS; den Herder et al. 2001), and we also examined
 the light curves
of the EPIC pn and MOS cameras. The calibrated energy
range of the EPIC cameras is 0.15-12 keV for the pn, and
0.3-12 keV for the MOS. The RGS wavelength range is 5-38
\AA, corresponding to the 0.33-2.5 keV energy range.
We performed the data extraction and analysis with XMM-SAS
 (XMM Science Analysis System) version 19.1, see
 \url{https://xmm-tools.cosmos.esa.int/external/xmm_user_support/documentation/sas_usg/USG.pdf}.
We extracted the RGS 1st order  spectra with the
XMM-SAS task rgsproc and combined the RGS1 and RGS2
 spectra with the RGSCOMBINE task. Periods of high background were
rejected.
 The total count rate was 2.446$\pm$0.005 cts s$^{-1}$,
 and the total  integrated flux we
 measured was 1.535 $\times 10^{-10}$ erg cm$^{-2}$ s$^{-1}$.

 Figure 5 shows a comparison between the {\sl Chandra}
 spectrum of 27 August 2021
 and the XMM-Newton spectrum obtained 3 days later in the 5--14 \AA \
 range in which the sensitivity is best matched. 
We observed a softening of
 the spectrum, with moderately decreased flux in lines formed at higher energy 
 and increased flux in lines formed at lower energy.
 As the EPIC-pn light curve in Figure 6 and the RGS light curve in the lower
 panel in the bottom plot on the right in Figure 7 show, there was
 moderate variability with some decrease in count rate over
 52 ks of ``clean'' observation. Particularly notable was a reduced flux
 in lines of oxygen, nitrogen and carbon between 18 and 35 
 \AA. Thus, we fitted the spectra of two separate intervals,
 indicated in blue and in orange in Figure 7. Strong lines are also
 marked for the first spectrum, and the many transitions are 
 indicated in the bottom panel. 
 In the first time interval, the total count rate was 2.514$\pm$0.008  cts s$^{-1}$,
 with an integrated flux of 1.584 $\times 10^{-10}$ erg cm$^{-2}$ s$^{-1}$.
 In the second interval the total
 count rate in the 5-35 \AA \
 wavelength range was 2.395$\pm$0.009 cts s$^{-1}$,
 with an integrated flux of 1.508  $\times 10^{-10}$ erg cm$^{-2}$ s$^{-1}$.

 We note that the G ratios of the spectrum integrated over the 
 whole spectrum are only around 1 for Si XIII and Mg XI (consistent with collisional
 ionization equilibrium), 
 while the R ratios for these two triplets,
 respectively, are 4.02$\pm$0.22 and 2.05$\pm$0.49, the latter implying
 an electron density n$_{e} \leq 10^{12}$ cm$^{-3}$, which is consistent 
 with the HETG result.
\subsection{Model fitting}
 Table 3 shows the parameters of the fit to the spectra, 
 observed in the two time intervals. At first, we performed the  
 fit with two plasma component as for the {\it Chandra} fit above.
 Such fits can be obtained with two BVAPEC components
 at $\simeq$0.1 keV and $\simeq$0.7 keV, respectively, yielding
 a value of $\chi$/d.o.f of 1.3. 
 However, fit the spectra with two
 components does not match the EPIC broad band spectra, which
 show a broad iron feature at $\simeq$6.7 keV (see Fig. 8) and
 require a third component at higher temperature. 
 The EPIC instruments cover a larger energy
 range than the RGS (0.15-12 keV for the pn
 and 0.3-12 keV for the two MOS), and fitting their spectra
 requires a thermal component in the 2.0-2.5 keV range,
 close to what we found for the HETG spectrum of 3 days earlier. 

 A joint fit with any of the three EPIC instruments could
 only be obtained with $\chi^2\geq$2.0:  the simultaneous 
 fit could not
 be done more rigorously, and we  assume that this is mainly 
because of the variability during the exposure. Moreover,
 we also have to factor the fact that, in order to
 avoid possible pile-up and allow better timing studies, the EPIC pn and MOS-1 
 exposures were done with the small window and the EPIC MOS-2 in timing mode, not allowing
 a subtraction of a simultaneously observed background. 
However, to obtain a consistent
 result we fitted also the RGS spectra (separately) with 
an additional component at a fixed plasma temperature
 of 2.4 keV, which appear to
 be necessary to fit the EPIC spectra
 and contributes to flux also in the RGS energy range. The spectrum of the second time interval
 has lower flux at low energy, so by using three components
 the uncertainty on the one at lower temperature became very large, thus
 we also fixed the value at a maximum of 130 eV, consistently with EPIC pn.
 We also ``froze'' one of the two values of N(H) to what we found
 for the first interval, N(H)=4.6 $\times 10^{21}$ cm$^{-2}$.
 These fits yield a larger $\chi^2$ than the ones with only two
 components, but qualitatively the single emission lines are better fitted. 
 Partly because
 of the sensitivity of the RGS to a softer portion of the spectrum,
 and partly because of what appears to be actual cooling
 of the plasma within 3 days, measuring more lines at longer 
 wavelengths we found it necessary to increase the number
 of free parameters. We assumed that at least two 
components were absorbed differently, and the
 component at intermediate
 temperature has a different N(H) parameter. This is inspired by 
 detailed hydrodynamical calculations \citep{Orlando2009},
that indicate shocked ejecta and shocked circumstellar medium can have distinctly different temperatures and intrinsic absorption.
 We also assumed that the components have different velocity, contributing
 differently to line broadening. The apparent blue shift was small,
 so we left it fixed for all components.
 The model fits are not perfect, 
 in fact the values of the $\chi^2$/(degrees of freedom) parameters 
 for the best fit 
 is only 1.8 in the first interval and 1.6 in the second, and the total
 flux is underestimated by 7-8\% relative
 to the actual measurement.  These RGS model fits are shown with 
red solid lines in Figure 7, and Figure 9 presents the
 fit to the softest portion of the RGS
 spectrum. In addition, Fig. 8 shows how the EPIC-MOS spectrum
 is fitted with the same composite model. Although also some of the L-shell lines
 of iron in the RGS are modeled with insufficient flux, we attribute the  
 non-perfect fit mainly to lines that we could not fit longward of 20~\AA, shown in Figure 9: the flux of the N VI $\gamma$
 line, for example, is largely underestimated, the O VII He-like triplet is
 not well fitted (most
 notably the G ratio is larger than in other He-like triplets,
 with a prominent forbidden line),  
   and we also could not fit well the  N VI He-like r line
 and a C VI $\gamma$ line. We attribute this to a complexity
 that may require different abundances for the three components, but
 we did not want to introduce a very large number of free parameters,
 since the fit to a high resolution spectrum with
 ``simple'' physical models gives important indications, but more accurate
 results may require detailed hydrodynamical models. 
\begin{table}
\caption{Parameters for the two RGS fits described in the text
 for the August 30 observation  (on day 21 after optical
 maximum)i split in two intervals, both fits including 
 three BVAPEC components. The parameters
 have the same meaning
 as in Table 1. Only the abundances that differ from solar
 within the statistical uncertainty are given. We assumed there are 
 two different values of N(H) and broadening velocity for each component.
}
\begin{center}
\begin{tabular}{ccc}
\hline
 & Interval 1 & Interval 2  \\
\hline
 $\chi^2$/d.o.f. & 1.8   & 1.6  \\
 N(H)$_{1,3}$ $\times 10^{21}$ cm$^{-2}$ & 6.1$\pm1.3$ & 6.4$\pm$1.0 \\
 N(H)$_{2}$ $\times 10^{21}$ cm$^{-2}$ & 4.6$\pm0.2$ & 4.6 (fixed)  \\
 kT$_{1}$ (eV) & 91$\pm$19 & 128$^{+300}_{-8}$     \\
 kT$_{2}$ (keV) &  0.64$\pm$0.06 & 0.64$\pm$0.01 \\
 kT$_{3}$ (keV) &  2.4 (fixed)   & 2.4 (fixed) \\
 EM$_{1} \times (d(Kpc)/2.4)^2$ & (1.298$\pm0.24) \times 10^{57}$ &
 (1.263$\pm0.125) \times 10^{57}$  \\
 EM$_{2} \times (d(Kpc)/2.4)^2$  & (1.056$\pm0.11) \times 10^{58}$ & (1.510$\pm0.17) \times 10^{57}$
   \\
 EM$_{3} \times (d(Kpc)/2.4)^2$ & (5.267$\pm 0.682) \times 10^{57}$   & (4.865$\pm 0.681) \times 10^{57}$  \\ 
 F$_a$ 10$^{-10}$ erg cm$^{-2}$ s$^{-1}$ & 1.52$\pm$0.26 & 1.39$\pm$0.18  \\
 F$_{a,1}$ 10$^{-12}$ erg cm$^{-2}$ s$^{-1}$ & 2.97$^{+1.17}_{-0.86}$ & 1.75$\pm$0.28  \\
 F$_{a,2}$ 10$^{-11}$ erg cm$^{-2}$  s$^{-1}$ & 10.40$\pm$1.38 & 9.96$\pm$1.15 \\
 F$_{a,3}$ 10$^{-11}$ erg cm$^{-2}$  s$^{-1}$ & 4.51$\pm$0.58 & 3.79$\pm$0.53 \\
 F$_u$ 10$^{-10}$ erg cm$^{-2}$ s$^{-1}$ & 6.38  & 7.49 \\ 
 C/C$_\odot$          & 7.95$^{+3.06}_{-2.56}$ & 0.1 (min.)    \\
 N/N$_\odot$          & 40.2$^{+6.6}_{-5.3}$  & 23$\pm$3 \\ 
 O/O$_\odot$          & 2.97$^{+0.43}_{-0.35}$ &  2.27$^{+0.22}_{-0.20}$ \\
 Ne/Ne$_\odot$        &  2.18$_{-0.25}^{+0.35}$  & 1.49$_{-0.14}^{+0.15}$ \\
 Mg/Mg$_\odot$        & 2.34$^{+0.32}_{-0.26}$ &  1.62$ ^{+0.16}_{-0.14}$ \\
 Al/Al$_\odot$        & 2.37$^{+0.65}_{-0.58}$ & 1.56$\pm$0.39 \\
 Si/Si$_\odot$        & 1.86$^{+0.25}_{-0.20}$ & 1.27$\pm$0.12 \\
 Fe/Fe$_\odot$        & 0.45$\pm$0.06 & 0.35$\pm$0.03 \\
 v$_{bs}$ km s$^{-1}$ & 474$\pm$24  & 525$\pm$24  \\
 v$_{br,1}$ km s$^{-1}$ & 1473$^{+144}_{-137}$ & 1505$^{+148}_{-153}$ \\
 v$_{br,2}$ km s$^{-1}$ & 759$\pm$36 & 734$\pm/$35  \\
 v$_{br,3}$ km s$^{-1}$ & 2125$^{+324}_{-327}$ & 2040$^{+327}_{-297}$ \\
\hline
\end{tabular}
\end{center}
\end{table}
\begin{figure}
\begin{center}
\includegraphics[width=130mm]{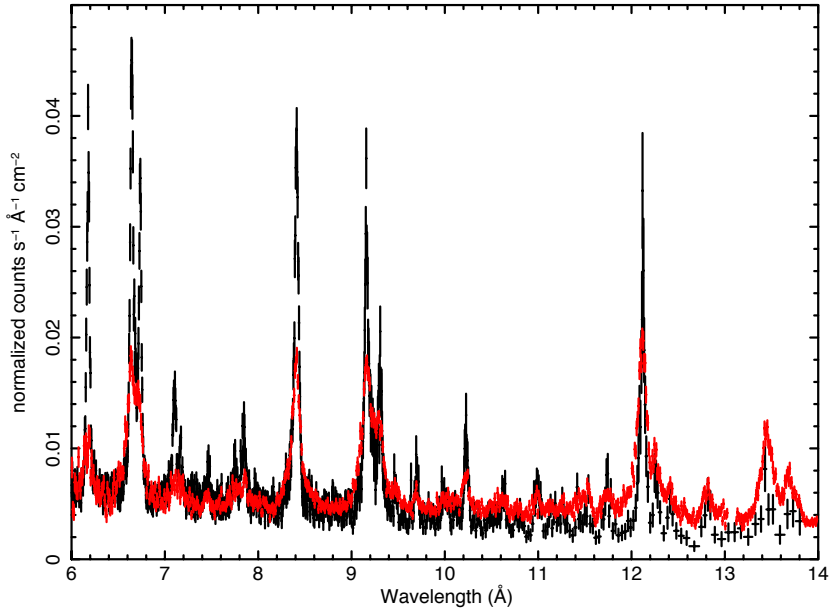}
\end{center}
\caption{Comparison of the Chandra MEG spectrum (in black) 
observed on 27 August 2021 
 and the XMM-Newton RGS spectrum (in red) observed on 30-31 August 2021 in the
 5--14 \AA \ wavelength range.}
\end{figure}
\begin{figure}
\begin{center}
\includegraphics[width=130mm]{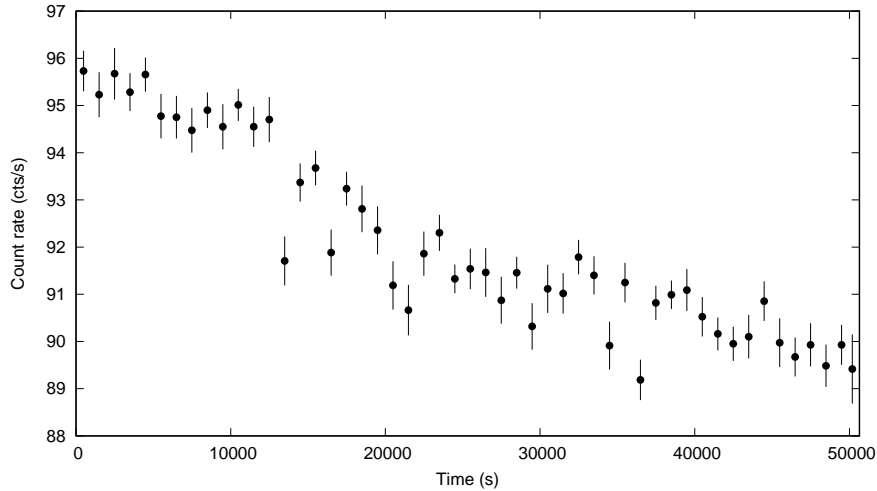}
\end{center}
\caption{The 0.15--10 keV lightcurve measured with EPIC pn,
 binned with bins of 1000 s.}
\end{figure}
\begin{figure}
\begin{center}
\includegraphics[width=110mm]{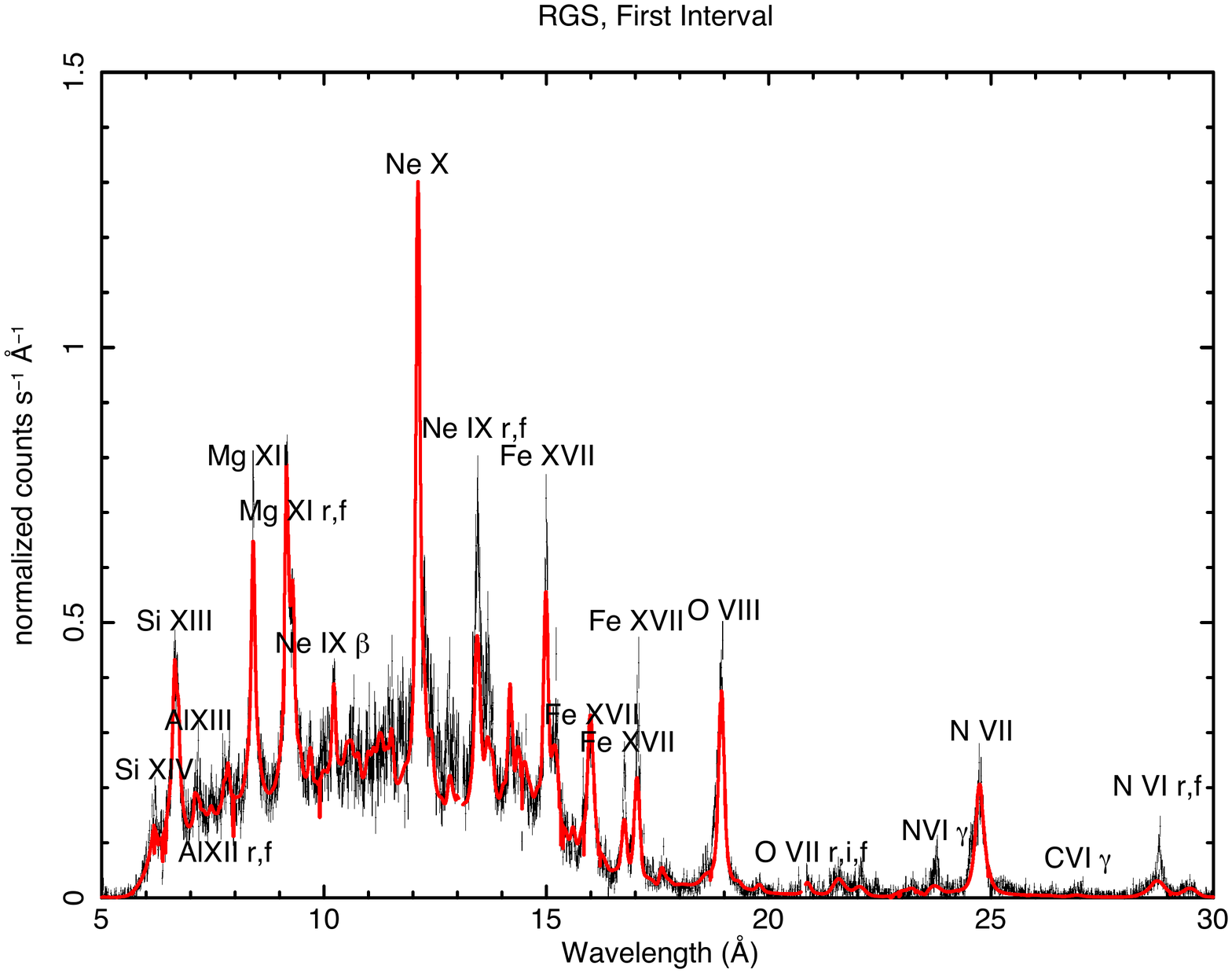}
\includegraphics[width=110mm]{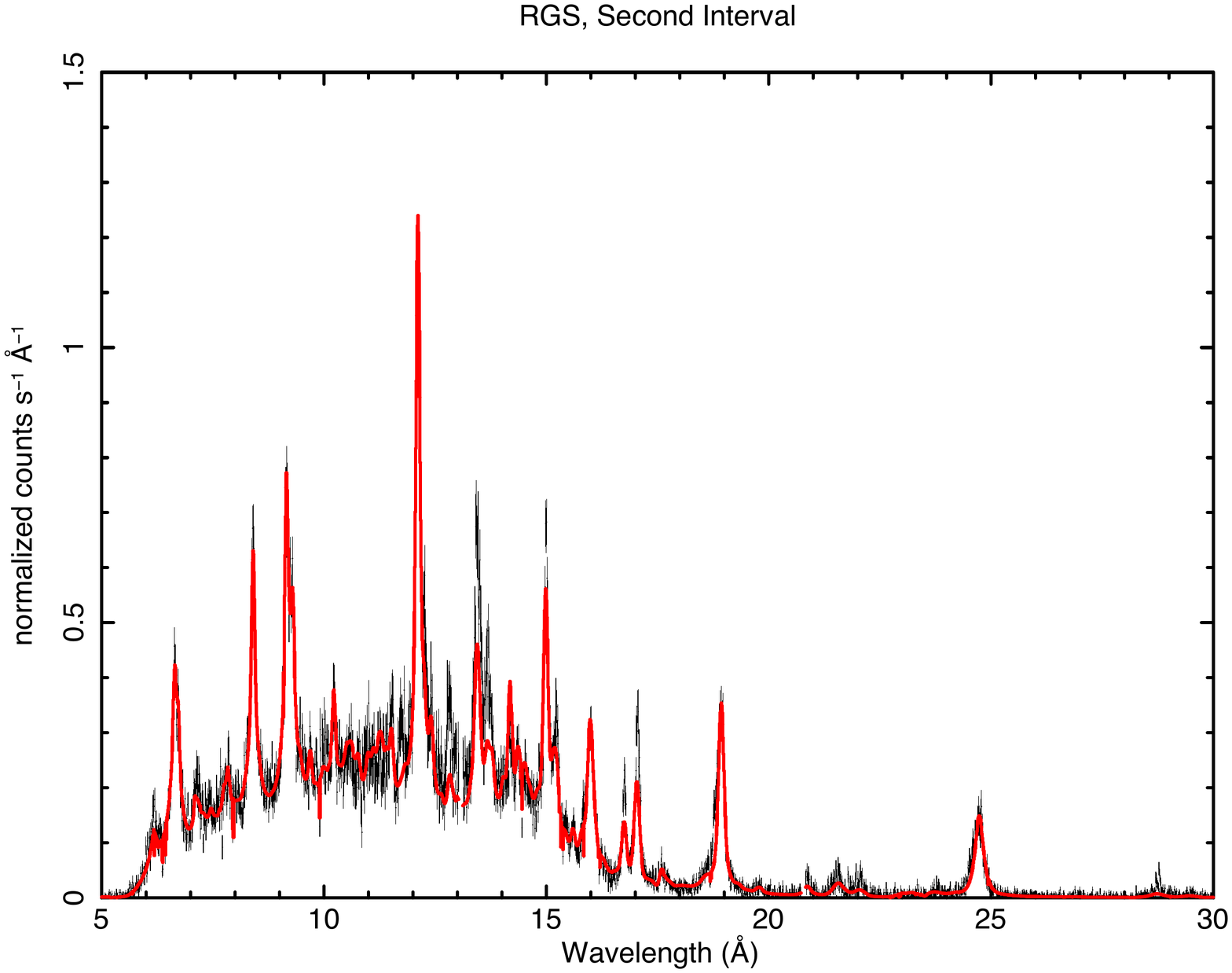}
\includegraphics[width=95mm]{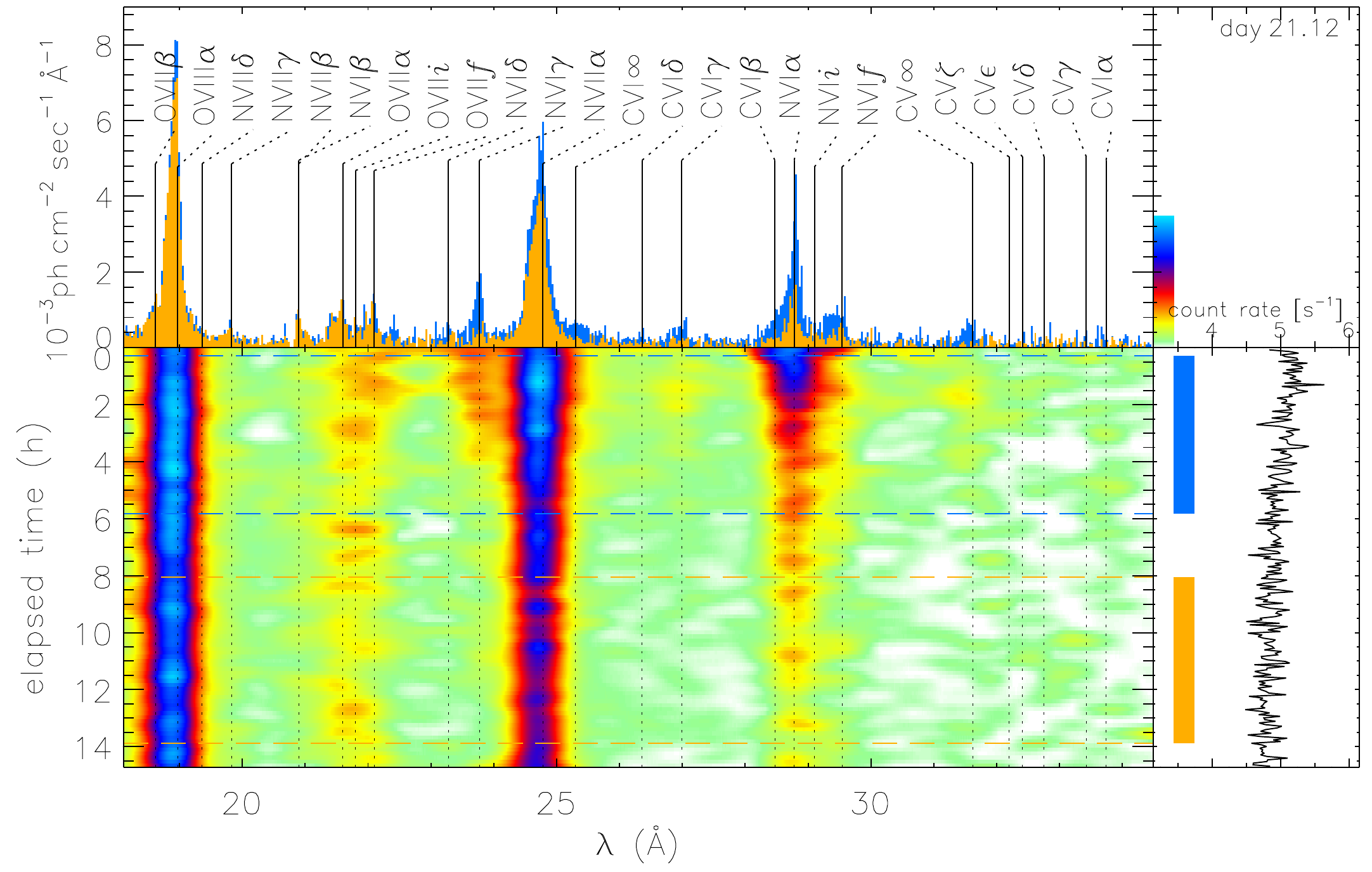}
\end{center}
\caption{The spectrum in the top and middle panels were extracted 
 from the time intervals marked in
 the blue and orange interval, respectively, in the lightcurve
 in the right lower panel. The top panel
 of this figure shows the comparison between the soft spectrum in the blue
 and orange interval, while the lower left panel shows the variation in
 time with the color map shown in the top right panel. The red solid lines
 are the models in the first column of Table 2 for the first interval,
 and the model in the third column of Table 2 for the second interval.}
\end{figure}
\begin{figure}
\begin{center}
\includegraphics[width=110mm]{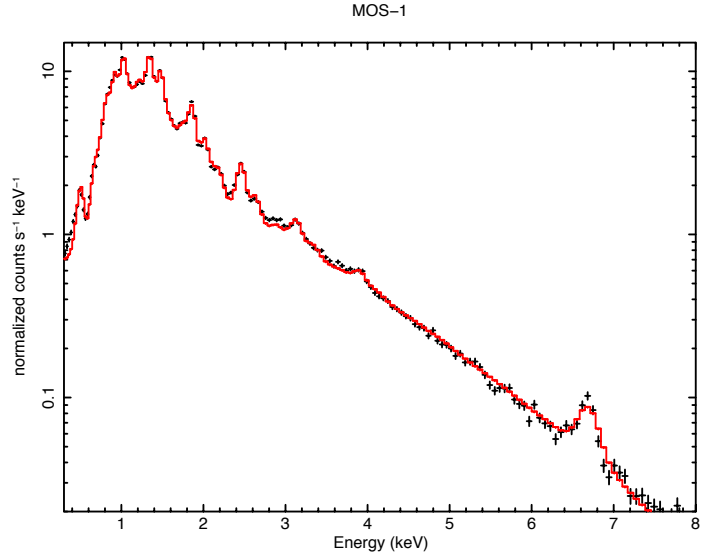}
\end{center}
\caption{The averaged MOS-1 spectrum over the whole
 XMM-Newton exposure, in logarithmic scale, and in red a model 
 with  three thermal components like the one in 
 Table 3. 
 The iron feature at $\simeq$6.7 keV is detected with low count rate, but 
 it is clearly present during the whole exposure time,
 and indicates the presence of the hottest component, even if its flux has decreased since day 18.
 }
\end{figure} 
\begin{figure}
\begin{center}
\includegraphics[width=110mm]{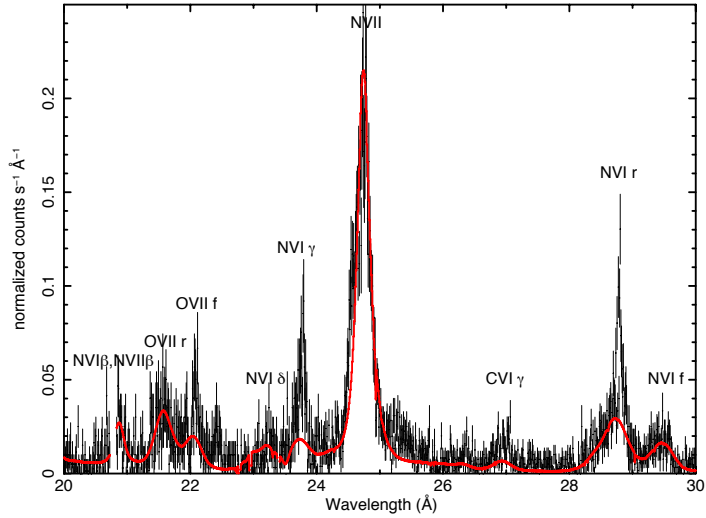}
\end{center}
\caption{The spectra of the first interval in
 the 20--30 \AA \ range, and the fit with the 
 model in Table 3 (red solid lines). The third component at lower
 temperature is necessary for the fit in this range, but it
 was not evident in the higher HETG energy range 3 days earlier. 
 }
\end{figure}
\begin{figure}
\begin{center}
\includegraphics[width=140mm]{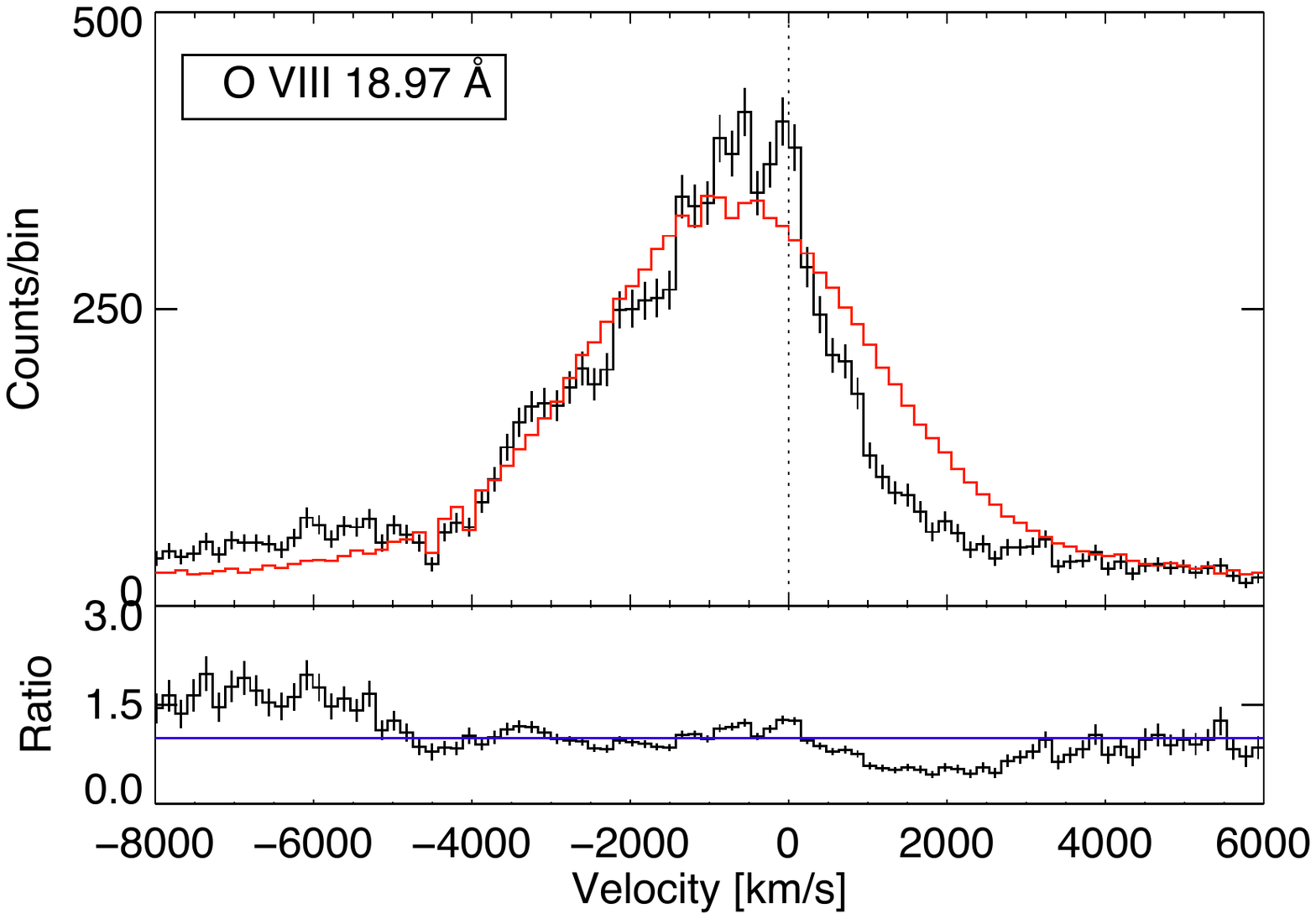}
\end{center}
\caption{The O VIII H-alpha line observed with the RGS,
 shown in velocity space with 0 at rest wavelength
 of 18.9761 \AA, and the fit with a Gaussian. The asymmetry, which we attribute
 to differential absorption in the receding outflow and in the one towards us, is
 more evident for lines with low ionization potential.
 }
\end{figure}
\subsection{Line fluxes and Line profiles.}
 Table 2 reports also the wavelength, broadening velocity and flux measured
 for the RGS, with the same Gaussian fits used for the HETG. 
  Like in the HETG case, the G ratio
 (1.68$\pm$0.52 for Mg XI, 1.15 for Si XIII) indicates a thermal plasma
 in collisional ionization equilibrium, without contribution of photoionization.

 The lower energy range of the RGS, and most likely also the fact that the
 ejecta have already expanded a bit more, allow to assess that
 the blue-shift effect is mainly due to differential absorption,
 as we see in Fig. 9. Note that fit to this O VIII line is not included in Table 2,
 because we found it differed too much from a Gaussian profile.
\section{Comparison with 2006 Observations}
{\sl Chandra} HETG spectra were obtained in 2006 earlier in the outburst,
 with the exposure
 starting 13.81 days after the outburst, about a week earlier than in 2021.
 The count rates were
 2.950$\pm$0.024 cts s$^{-1}$ in the HEG
 and 1.271$\pm$0.014 cts s$^{-1}$ in the MEG,
 and the total flux was significantly higher than in 2021 on day
 20: 3.3 $\times 10^{-10}$ erg s$^{-1}$ cm$^{-2}$ 
 measured with the MEG and 4.1 $\times 10^{-10}$ erg s$^{-1}$ cm$^{-2}$ 
 with the HEG. Figure 11 shows these spectra
 and a fit with the same model adopted for 2021.
 A comparison of the two spectra shows that the continuum in 2006 was
 higher and the emission lines were stronger, but the same lines 
 were observed. The Fe XXV line at 1.778 \AA \
 was stronger compared to other lines, indicating a higher plasma
 temperature. \citet{Nelson2008}
 fitted the spectrum together with one from the RGS observed on the
 same day, reaching longer wavelengths,
 and obtained an optimal fit
 with 4 thermal plasma components at different temperature, 16.84, 2.31, 0.92
 and 0.64 keV and NH)=1.2 $\times 10^{22}$ cm$^{-2}$.     
 However, we find
 that with only two components of APEC plasma at 4.84 keV and 0.76 keV,
 respectively, and N(H)=5.6 $\times 10^{21}$ cm$^{-2}$, we
 can fit the 2006 HEG and
 MEG spectra in the 1.2--15 keV range with a reduced $\chi^2$=1.18. 
 Most of the emission lines are well fitted, although in the soft range some flux
 is underestimated. This fit indicates that most abundances are slightly
 enhanced with respect to solar, except that of iron that turns out to be
 only about half of the solar value, like in 2021, probably indicating mixing 
 with the red giant wind. 

   Next, we examined the 2006 RGS spectrum observed 
 in an exposure starting exactly 26.13 after maximum, to
 compare it with our 2021 RGS spectrum observed on day 21.  
  We re-examined  the flare claimed by \citet{Nelson2008}, shown in their
 EPIC-pn light curve of the timing-mode
 exposure.  We compared
 the light curve measured with the RGS gratings with that of an event file
 accumulated only from the exterior part of the EPIC-MOS-2 camera,
 without collecting photons from the source, to analyze the
 background, and found that the variability observed in the RGS 
 and EPIC-pn was due to 
 a background flare, and cannot be attributed to the source. We ``cleaned''
 the March 2006 spectrum excluding intervals of high background: 
 the ``new lines'' are not appearing suddenly
 during the exposure, but are always present when examining
 only intervals in which the background flare is excluded.
  We combined the RGS1 and RGS2 count rate of the 1st order, measuring
 a count rate of 2.819$\pm$0.014 cts s$^{-1}$, only a little higher
 than in 2021.
 While the spectrum below 20~\AA \ is a little more luminous and
 the line ratios are almost the same,  Figure 11 
 shows that instead the spectrum in the 20--30 \AA \ region differs considerably.
 Most of the unidentified lines of 2006 are not present in the 2021
 spectrum, but we note two important facts: a) The common features
 are identified without assuming a large blueshift for the 2006
 spectrum (a blueshift velocity of 8000 km s$^{-1}$ was proposed
 by \citet{Nelson2008} in order to identify some
 of the lines), b) The emission lines
 that are not common to the 2021 spectrum presented here did indeed appear
 in 2021 some time later in the outburst, in a spectrum
 measured in 2021 with the RGS about two weeks later, and
 will be described in a paper
 in preparation led by coauthor Ness.
\begin{figure}
\begin{center}
\includegraphics[width=130mm]{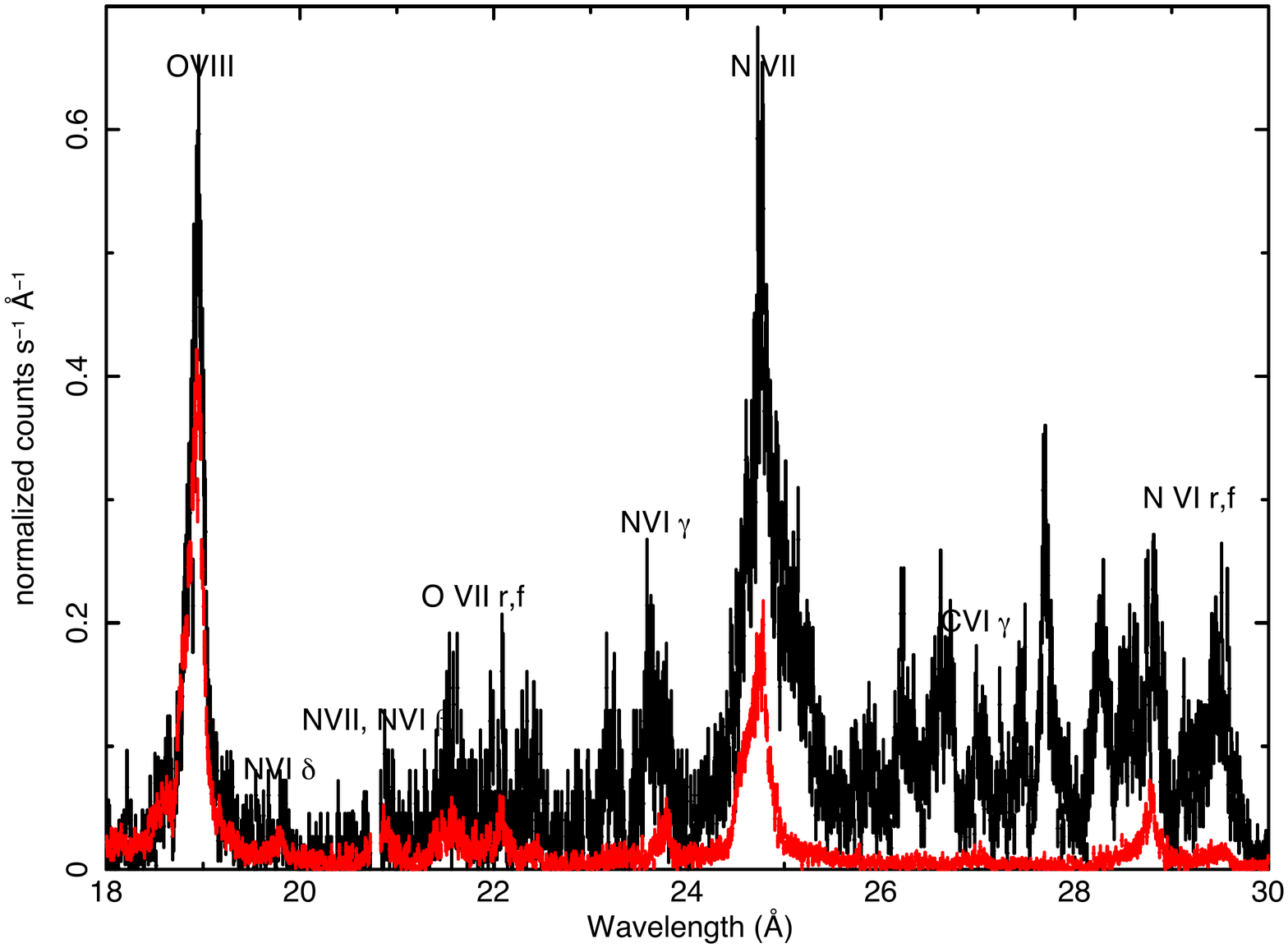}
\end{center}
\caption{The RGS spectrum measured on 2006 March 10, 26 days after
 the optical maximum, is plotted in black and compared in the 20-30 \AA \ range with
 the RGS spectrum observed on 2021 August 30, 21 days after the optical
 maximum, plotted in red.}
\end{figure} 
\section{Discussion}
 The high spectral resolution spectra we obtained cover relatively 
 narrow ranges, but they are rich in many emission lines that constrain
 the plasma temperatures. 
 Adding information obtained in broad band in  wider
 energy ranges (with {\sl NICER} less than a day
 earlier for the HETG, and with the simultaneous EPIC
 spectra for the RGS), we were able to better constrain
 the value of the absorbing
 column density N(H) and plasma temperatures
 and to some extent also the abundances, especially
 that of iron. We find that, in the fits to 
 the Swift X-Ray Telescope (XRT) monitoring \citep[][their Fig. 11]{Page2022}, 
 the value of N(H) (about 10$^{22}$ cm$^{-2}$)
was overestimated for days 18 and 21 
 in the  two-thermal components fits of the XRT broad band spectra.
 The plasma temperatures of two components estimated with
 the Swift XRT are in the same ballpark 
 measured with the gratings, but the Swift XRT fits miss the
 emergence of a much softer component at wavelengths
 longward of 23 \AA, and is evident in the {\sl XMM-Newton}
 RGS spectrum. This component became important in
 the following evolution, because in broad
 band spectra it is easily confused with the emergence of the central
 supersoft source, generating confusion in estimating the
 amount of ejected mass and/or some crucial physical parameters.  

  We confirm that the high resolution spectra reveal a very low
 filling factor of the ejecta and 
 the effect of differential intrinsic absorption in the material flowing away from
 our line of sight and towards it. The iron abundance in the fits, driven 
 mostly by the L-shell iron lines, is lower than solar, at most 
 half the solar value, but possibly even quite lower. 
 The RGS spectrum of day 21 indicates also high nitrogen abundance, 
possibly around 40 times solar in the spectrum of the first part
 of the exposure. This is consistent with
 the optical spectra of \citet{Pandey2022}. It is thus likely that at this point
 CNO processed ejected material was mixing with the circumstellar red giant
 wind. However, iron remained depleted. In the optical nebular spectrum
 of outburst day 201 after the 1985 outburst maximum, \citep{Contini1985}
 found 10 times enhanced nitrogen, the ratio of carbon to nitrogen 100 times
 the solar value, and iron abundance only 17\% the solar value, so
 these results seem to be consistent throughout outburst cycles.

 If we interpret the line broadening as due to the velocity of the emitting
 medium, it is interesting to note that on day 18 this velocity is
 quite lower than that derived in the optical spectra, but it appears to
 have increased on day 21 in the lines formed at lower and at higher
 energy. \citet{Aydi2020} noticed that in all
 novae  optical spectra a fast 
 velocity component follows a slower one, and often there is evidence
 of two or multiple mass ejection episodes. The varying 
 velocity estimates of the X-ray emitting ejecta of RS Oph and the discrepancy
 with the maximum velocity observed in its optical spectra
 are consisting with this complex scenario and indicate that the shocked
 material is associated with the slower outflow
 episodes.

\section{Conclusions}
 We can summarise the main conclusions as follows:

1. Assuming the distance estimate given by \citet{Rupen2008},
 the precisely measured flux on day 18 implies an X-ray 
luminosity 1.738 $\times 10^{35}$ erg s$^{-1}$ in the 0.4-10 keV range
and 1.023  $\times 10^{35}$ erg s$^{-1}$ in the 0.33-2.5 keV range
 on day 21. The best fits indicate absolute luminosity of about 
 3.9 $\times 10^{35}$ erg s$^{-1}$ on day 18  and 5.2 $\times 10^{35}$ 
erg s$^{-1}$ on day 21, confirming
 that shocks in the outflows of symbiotic novae produce
 very luminous X-ray sources.

2.  18 days after the optical maximum, the HETG X-ray high
 resolution spectra of RS Oph in the 2021 outburst are fitted with 
 a thermal plasma with at least two components at different temperature.
 Observing in the lower energy range with the {\sl XMM-Newton} RGS 
 on day 21, we measured emission lines produced in a 
 third thermal component, at about 0.1 keV. Although
 the HETG was not sensitive in the softest range, the {\sl NICER} spectra in
 the 0.2-12 keV range on day 18 pose low flux limits on this soft 
 component on day 19 (Orio et al 2022, in preparation), so we suggest 
 that this component
  was only emerging between the two dates of our observations,
 spaced 3 days apart.
 We anticipate that comparison with another {\sl XMM-Newton}
 RGS spectrum obtained at a later outburst phase
 indicates that this soft thermal component appears to become more luminous with
 time between the third and the fifth week after the outburst
 (Ness et al. 2022, in preparation).

3. An interesting parameter is the column density, proving 
 intrinsic absorption of the ejecta. In fact, the interstellar column density
 was estimated to be 2.4$\pm$0.6 $\times 10^{21}$ cm$^{-2}$ from
 H I absorption by \citet{Hjellming1986}. 
 However, the values obtained in our
 fits indicate much lower intrinsic ``wind absorption'' 
 than estimated by \citet{Page2022} with the {\sl Swift} data.
 On day 18, these authors estimate an additional N(H) of 1.50-1.54
 $\times 10^{22}$ cm$^{-2}$, which is so high to be ruled out even 
 by the fits to the HETG alone (without resorting to the NICER data).
 These authors also  estimate a wind absorption still as high as 
 N(H)=1.37 $\times 10^{22}$ cm$^{-2}$ on day 21, and
 conclude that the outflowing material had exactly
 the same characteristics in 2021.  We would like to point out that 
 it is difficult and not always 
 reliable to draw firm conclusions only from the broad band data,
 while a combination of broad band and high resolution X-ray spectra
 constrains the value of the column density much more accurately.

4.  An important advantage of our high resolution spectra is the precise
 determination of the absorbed flux. This cannot be done with
 broad band spectra, which rely on modeling to give a flux estimate.
The flux measured with the {\sl XMM-Newton} RGS on day 26 after  the
 2006 outburst maximum
 was 1.6 $\times 10^{-10}$ erg cm$^{-2}$ s$^{-1}$,  still 
 a little higher than what we measured 
 earlier in post-outburst phase,
 on day 21 in 2021. Most likely, this indicates faster cooling
 of the plasma in the 2021 outburst.
  The indications of lower intrinsic absorption and more rapidly decreasing
 flux in 2021 compared to the 2006 data of \citet{Nelson2008}
 indicate that the ejected mass was larger in 2006. 

5.  Another very important caveat is derived by the comparison with the $\gamma$-ray emission
 in the 0.2-5 GeV range of the Fermi LAT \citep{Cheung2022}. Around
 day 20 post-optical maximum, the flux measured with Fermi was
 about 10$^{-7}$ erg cm$^{-2}$ s$^{-1}$. With the gratings in the X-ray range,
 we are able to measure the X-ray flux  precisely,
 and on day 18 we only measured 2.6 $\times 10^{-10}$ erg cm$^{-2}$ s$^{-1}$
 in the 0.5-10 keV range (see the Chandra data), three orders of
 magnitude less than in the Fermi range. Not
 only the shock energy loss by particles appears to exceed the X-ray emission from the hot gas by such a large
 amount, but also the plasma temperature of the detectable
 emission is
 much lower than predicted by \citet{Vurm2018}, constraining the models
 of $\gamma$-ray emission and most likely implying that the site of the shocks
we observed is {\it not} the same site of the shock whose consequences
 were observed  with Fermi. The proposed explanation for
 the absence of simultaneous $\gamma$-ray and X-ray
 flux consistent with each other,  in several papers is
 that the X-rays associated with the $\gamma$-ray flux are 
 affected by too high absorption to be detected \citep[e.g.][]{Vurm2018}, but
 our spectra rule out such high column density.
 This result is new and important, because only during
 the RS Oph outburst $\gamma$-ray and X-ray flux were measured simultaneously
 in a  nova.
 
6.  The only abundances that are quite clearly constrained are those of
 nitrogen (enhanced with respect to solar values) and of iron (which
 is depleted), consistently with results inferred after a previous
 outburst \citep{Contini1985}. However, there is some evidence
 that the abundances in the two components may not be the same, and most 
 elements, except iron, may be enhanced in the less cool component,
 and closer to solar values in the hotter one. Moreover, 
 the fit on day 19 can be obtained with abundances that are close to 
 solar, while on day 22 a better fit is obtained with enhanced
 values, perhaps indicating ongoing mixing with the ejecta (there
 are. however, quite large uncertainties). 
 The spectrum on day 21 is better fitted with different absorbing
 column density for the two components, with values 
 that are still significantly higher than the
 interstellar value, although only about half of the estimate
 of \citet{Page2022}. 

7. We also note that on day 21, we did not
 measure  yet many of the - partially unidentified - emission lines in
 the soft range measured on day 26 of the outburst
 in 2006. During this {\sl XMM-Newton} exposure, we also observed a 
 decrease in count rate, due mainly to decreasing
 flux in the emission lines in the soft region. 
  Separate fits to the spectrum observed in
 the first part of the day 21 exposure and in the second half are more
 consistent with variable flux of
 the softer component than with variable absorption in the ejecta. 
 However, we anticipate that the emission lines in this region
 later on in the outburst become stronger, more lines were
 measurable and the flux of the softer component increased:
 this development is described in paper describing
 with subsequent exposures with the RGS
 gratings (Ness et al. 2022, in preparation) and with
 {\sl NICER} (Orio et al. 2022, in preparation).

8. Finally, these data and their analysis,  including
 the line profiles, the column density estimate and the low velocity
 associated with the X-ray emitting outflow, 
 should allow refining sophisticated
 models of the dynamics of the ejecta like those of \citet[][]{Orlando2009} 
 for the 2006 outburst.

 A second article with other high
 spectral resolution X-ray grating
 spectra of RS Oph is in the works, led by coauthor Ness, and will
 cover the initial emergence of the supersoft X-ray source.

\facilities{ Chandra - XMM-Newton
}
\software{CIAO v4.14.0 \citep{Fruscione2006}, XSPEC v6.30.6
 \citep{Arnaud1996, Arnaud2022}, GRPPHA \citep[see][and references therein]{Dorman2001},
  SAS v19.1 \url{https://xmm-tools.cosmos.esa.int/external/xmm_user_support/documentation/sas_usg/USG.pdf}
}

\bibliography{rsbiblio.bib}{}
\bibliographystyle{aasjournal}


\begin{acknowledgements}
 MO acknowledges useful and interesting conversations
 with Joe Cassinelli and Tommy Nelson. The immediate comparison with the NICER 
 data was possible thanks to Keith Gendreau. 
MO was supported by NASA and NASA-Smithsonian awards for
 research with XMM-Newton and Chandra, respectively.
GJML is member of the CIC-CONICET (Argentina) and acknowledges support from grant ANPCYT-PICT 0901/2017. JM was financed by the
 Polish National Science Centre grant 2017/27/B/ST9/01940.
\end{acknowledgements}
\end{document}